\documentclass[journal]{IEEEtran}

\usepackage{mathrsfs}
\usepackage{amssymb}
\usepackage{bbm}
\usepackage{epsfig}
\usepackage{amsmath}
\usepackage{color}
\usepackage{amsfonts}
\usepackage{dsfont}
\usepackage[english]{babel}
\usepackage[mathscr]{eucal}
\usepackage{graphicx}
\usepackage{amsbsy}
\usepackage{amsopn}
\usepackage{amstext}
\usepackage{latexsym}
\usepackage{color,multicol,makeidx}
\usepackage{CJK}
\usepackage{indentfirst}
\usepackage{bm}
\usepackage{flushend}
\usepackage{balance}
\usepackage{float}
\usepackage{ccaption}
\usepackage[caption=false,font=normalsize,labelfont=sf,textfont=sf]{subfig}
\usepackage{multirow}

\newtheorem{them}{Theorem}[section]
\newtheorem{defn}{Definition}[section]

\newtheorem{lem}{Lemma}[section]

\newtheorem{rem}{Remark}[section]
\newtheorem{exa}{Example}[section]
\ifCLASSINFOpdf
\else
\fi
\begin{document}
\title{Nash Equilibrium Computation in Subnetwork Zero-Sum Games with Switching Communications}

\author{Youcheng~Lou, Yiguang~Hong,~\IEEEmembership{Senior~Member,~IEEE,} Lihua~Xie,~\IEEEmembership{Fellow,~IEEE,} Guodong~Shi and~
Karl~Henrik~Johansson,~\IEEEmembership{Fellow,~IEEE}
\thanks{
Recommended by Associate Editor D. Bauso.
 This work is supported by the NNSF of
China under Grant 71401163, 61333001, Beijing Natural Science Foundation under Grant 4152057, National Research Foundation of Singapore under grant NRF-CRP8-2011-03, Knut and Alice Wallenberg Foundation, and the Swedish Research Council.}
\thanks{Y. Lou and Y. Hong are with
Academy of Mathematics and Systems Science, Chinese Academy of Sciences, Beijing 100190 China
 (e-mail: louyoucheng@amss.ac.cn, yghong@iss.ac.cn)}
\thanks{L. Xie is with School of Electrical and Electronic Engineering,
Nanyang Technological University, Singapore, 639798
       (email: elhxie@ntu.edu.sg)}
\thanks{G. Shi is with College of Engineering and Computer Science, The Australian National University, Canberra, ACT 0200 Australia (email: guodong.shi@anu.edu.au)}
\thanks{K. H. Johansson is with ACCESS
Linnaeus Centre, School of Electrical Engineering, Royal Institute
of Technology, Stockholm 10044 Sweden (e-mail: kallej@ee.kth.se)}
}

\markboth{IEEE Transactions on Automatic Control}
{Y. Lou \MakeLowercase{\textit{et al.}}: Nash Equilibrium Computation in Subnetwork
Zero-Sum Games}

\maketitle

\begin{abstract}
In this paper, we investigate a distributed Nash equilibrium
computation problem for a time-varying multi-agent network
consisting of two subnetworks, where the two subnetworks share the same objective
function. We first propose a subgradient-based
distributed algorithm with heterogeneous stepsizes to compute a Nash equilibrium of a zero-sum
game.  We then prove that the proposed algorithm can achieve a Nash equilibrium under uniformly
jointly strongly connected (UJSC) weight-balanced digraphs with
homogenous stepsizes. Moreover, we demonstrate that for weighted-unbalanced graphs a Nash equilibrium may not be achieved with homogenous
stepsizes unless certain conditions on the objective function hold. We show that there always exist heterogeneous
stepsizes for the proposed algorithm to guarantee that a Nash equilibrium can be achieved for UJSC digraphs. Finally,
in two standard weight-unbalanced cases, we verify the convergence to a Nash equilibrium by adaptively updating the stepsizes along with the arc weights in the proposed algorithm.
\end{abstract}


%
\IEEEpeerreviewmaketitle

\section{Introduction}

In recent years, distributed control and optimization of multi-agent
systems have drawn much research attention due to their broad applications
in various fields of science, engineering, computer science, and social science.
Various tasks including consensus, localization,
and convex optimization can be accomplished cooperatively for a group of autonomous
agents via distributed algorithm design and local information exchange \cite{cao1,xie1,rabbat,nedic1,nedic2,shi2,shi1,lou2}.

Distributed optimization has been widely investigated for agents to achieve a global optimization objective
by cooperating with each other \cite{nedic1,nedic2,shi2,shi1,lou2}.
Furthermore, distributed optimization algorithms in the presence of
adversaries have gained rapidly growing interest
\cite{boyd,Heg98,cor2,fri,sta}.  For instance,
a non-model based approach was proposed for seeking a Nash equilibrium of noncooperative
games in \cite{fri}, while distributed methods to compute Nash equilibria based on extreme-seeking technique were developed in \cite{sta}.
A distributed continuous-time set-valued dynamical system
solution to seek a Nash equilibrium of zero-sum games was first designed for undirected graphs
and then for weight-balanced directed graphs in \cite{cor2}.
It is worthwhile to mention that, in the special case of additively separable objective functions, the considered
distributed Nash equilibrium computation problem is equivalent to the well-known
distributed optimization problem: multiple agents cooperatively
minimize a sum of their own convex objective functions \cite{bj08,joh09,nedic1, nedic2,tou10, ram,zhu,chenj,cor14,elia}.

One main approach to distributed optimization is based on
subgradient algorithms with each node computing a
subgradient of its own objective function.
Distributed subgradient-based algorithms with constant and time-varying
stepsizes, respectively, were proposed in
\cite{nedic1,nedic2} with detailed convergence analysis.
A distributed iterative
algorithm that avoids choosing a diminishing stepsize was proposed
in \cite{elia}.
Both deterministic and randomized versions of distributed
projection-based protocols were studied in \cite{shi2,shi1,lou2}.

In existing works on distributed optimization, most of the results were obtained for switching
weight-balanced graphs because there
usually exists a common Lyapunov function to facilitate the convergence analysis in this case \cite{nedic1,nedic2,zhu,cor2,cor14}.
Sometimes, the weight-balance condition is hard to preserve in the case when the graph is time-varying and with communication delays \cite{tsi1}, and it may be quite restrictive and difficult to verify in a distributed setting.
However, in the case of
weight-unbalanced graphs, there may not exist a common (quadratic)
Lyapunov function or it may be very hard to construct one even
for simple consensus problems
\cite{tsi}, and hence, the convergence analysis of distributed problems
become extremely difficult. Recently, many efforts have been made to handle the weight unbalance problem, though very few results have been obtained on distributed optimization.
For instance, the effect of the Perron vector of the adjacency matrix on the optimal
convergence of distributed subgradient and dual averaging algorithms were investigated for a fixed weight-unbalanced graph in \cite{tsi3, sayed}. Some methods were developed for the unbalanced graph case such as the reweighting technique \cite{tsi3} (for a fixed graph with a known Perron vector) and the subgradient-push methods \cite{ned,nedic5} (where each node has to know its out-degree all the time).
To our knowledge, there are no theoretical results on distributed Nash equilibrium computation for switching weight-unbalanced graphs.

In this paper, we consider the distributed zero-sum game Nash equilibrium
computation problem proposed in \cite{cor2}, where a multi-agent network consisting of two subnetworks,
with one minimizing the objective function and the other maximizing it.
The agents play a zero-sum game. The agents in two different subnetworks play
antagonistic roles against each other, while the agents in the same
subnetwork cooperate.
The objective of the network is to achieve a Nash equilibrium via distributed computation based on local communications under time-varying connectivity.
The considered Nash equilibrium computation problem is motivated by power allocation problems \cite{cor2} and saddle point searching problems arising from Lagrangian dual optimization problems \cite{nedic,zhu,arrow,gol,durr,mai}. The contribution of this paper can be summarized as follows:
\begin{itemize}
\item We propose a
subgradient-based distributed algorithm to compute a saddle-point Nash equilibrium under time-varying graphs,
and show that our algorithm with
homogeneous stepsizes can achieve a Nash equilibrium
under uniformly jointly strongly connected (UJSC) weight-balanced
digraphs.

\item We further consider the weight-unbalanced
case, though most existing results on distributed optimization were obtained for
weight-balanced graphs, and show that distributed homogeneous-stepsize algorithms may
fail in the unbalanced case, even for the special case of identical subnetworks.

\item We propose
a heterogeneous stepsize rule and study how to
cooperatively find a Nash equilibrium in general weight-unbalanced cases. We find that,
for UJSC time-varying digraphs, there always exist
(heterogeneous) stepsizes to make the network achieve a Nash
equilibrium. Then we construct an adaptive algorithm to update the
stepsizes to achieve a Nash equilibrium in two standard cases: one with a common left eigenvector associated with eigenvalue one of adjacency matrices
and the other with periodically switching graphs.
\end{itemize}

The paper is organized as follows. Section II gives some preliminary knowledge, while Section III formulates the
distributed Nash equilibrium computation problem and proposes a novel
algorithm. Section IV provides the main results followed by
Section V that contains all the proofs of the results. Then Section VI provides
numerical simulations for illustration.  Finally, Section VII gives
some concluding remarks.

Notations: $|\cdot|$ denotes the Euclidean norm,
$\langle\cdot, \cdot\rangle$ the Euclidean inner
product and $\otimes$ the Kronecker product.
$\textbf{B}(z,\varepsilon)$
is a ball with $z$ the center and $\varepsilon>0$ the radius,
$\mathcal{S}^+_n=\{\mu|\mu_i>0,\sum^n_{i=1}\mu_i=1\}$ is the set of
all $n$-dimensional positive stochastic vectors.
$z'$ denotes the transpose of vector $z$,
$A_{ij}$ the $i$-th row and $j$-th column entry of matrix $A$ and
diag$\{c_1,\dots, c_n\}$ the diagonal matrix with diagonal elements $c_1,...,c_n$.
$\textbf{1}=(1,...,1)'$ is the vector of all ones with appropriate dimension.

\section{Preliminaries}

In this section, we give preliminaries on graph theory \cite{God},
convex analysis \cite{Roc}, and Nash equilibrium.

\subsection{Graph Theory}

A digraph (directed graph) $\bar {\mathcal{G}}=(\bar
{\mathcal{V}},\bar {\mathcal{E}})$ consists of a node set $\bar
{\mathcal{V}}=\{1,...,\bar n\}$ and an arc set $\bar
{\mathcal{E}}\subseteq \bar {\mathcal{V}} \times \bar
{\mathcal{V}}$. Associated with graph $\bar {\mathcal{G}}$, there is
a (weighted) adjacency matrix $\bar A=(\bar a_{ij})\in \mathbb{R}^{\bar
n\times \bar n}$ with nonnegative adjacency elements $\bar a_{ij}$,
which are positive if and only if $(j,i) \in \bar {\mathcal{E}}$.
Node $j$ is a neighbor of node $i$ if $(j,i) \in \bar
{\mathcal{E}}$.  Assume $(i,i)\in\bar {\mathcal{E}}$ for
$i=1,...,\bar n$.
A path in $\bar {\mathcal{G}}$ from $i_1$ to $i_p$ is an alternating
sequence $i_1e_1i_2e_2\cdots i_{p-1}e_{p-1}i_p$ of nodes $i_r, 1\leq
r\leq p$ and arcs $e_r=(i_r, i_{r+1})\in \bar {\mathcal{E}}, 1\leq
r\leq p-1$. $\bar {\mathcal{G}}$ is said to be bipartite if
$\bar {\mathcal{V}}$ can be partitioned into two disjoint parts
$\bar {\mathcal{V}}_1$ and $\bar {\mathcal{V}}_2$ such that $\bar
{\mathcal{E}}\subseteq\bigcup^2_{\ell=1}(\bar
{\mathcal{V}}_\ell\times\bar {\mathcal{V}}_{3-\ell})$.

Consider a multi-agent network $\Xi$ consisting of two subnetworks
$\Xi_1$ and $\Xi_2$ with respective $n_1$ and $n_2$ agents.  $\Xi$
is described by a digraph, denoted as
$\mathcal{G}=(\mathcal{V},\mathcal{E})$, which contains self-loops,
i.e., $(i,i)\in \mathcal{E}$ for each $i$. Here $\mathcal{G}$ can be
partitioned into three digraphs:
$\mathcal{G}_\ell=(\mathcal{V}_\ell,\mathcal{E}_\ell)$ with
$\mathcal{V}_\ell=\{\omega^\ell_1,...,\omega^\ell_{n_\ell}\},\;\ell=1,2$,
and a bipartite graph
$\mathcal{G}_{\bowtie}=(\mathcal{V},\mathcal{E}_{\bowtie})$, where
$\mathcal{V}=\mathcal{V}_1\bigcup \mathcal{V}_2$ and
$\mathcal{E}=\mathcal{E}_1\bigcup\mathcal{E}_2\bigcup\mathcal{E}_{\bowtie}$.
In other words, $\Xi_1$ and $\Xi_2$ are described by the two
digraphs, $\mathcal{G}_1$ and $\mathcal{G}_2$, respectively, and the
interconnection between $\Xi_1$ and $\Xi_2$ is described by
$\mathcal{G}_{\bowtie}$. Here $\mathcal{G}_{\bowtie}$ is
called bipartite without isolated nodes if, for any
$i\in\mathcal{V}_\ell$, there is at least one node
$j\in\mathcal{V}_{3-\ell}$ such that $(j,i)\in\mathcal{E}$ for
$\ell=1,2.$ Let $A_\ell$ denote the adjacency matrix of
$\mathcal{G}_\ell, \ell=1,2$. Digraph $\mathcal{G}_\ell$ is strongly connected if there is a path in $\mathcal{G}_\ell$ from
$i$ to $j$ for any pair node $i,j\in\mathcal{V}_\ell$. A node is
called a root node if there is at least a path from this node to
any other node.
In the sequel, we
still write $i\in\mathcal{V}_\ell$ instead of
$\omega^\ell_i\in\mathcal{V}_\ell,\; \ell=1,2$ for simplicity if
there is no confusion.

Let $A_\ell=(a_{ij},_{i,j\in \mathcal{V}_\ell})\in \mathbb{R}^{
n_\ell\times n_\ell}$ be the adjacency matrix of $\mathcal{G}_\ell$. Graph $\mathcal{G}_\ell$ is weight-balanced if
$\sum_{j\in \mathcal{V}_\ell}a_{ij}=\sum_{j\in
\mathcal{V}_\ell}a_{ji}$ for $i\in\mathcal{V}_\ell$; and
weight-unbalanced otherwise.

A vector is said to be stochastic if all its components are
nonnegative and the sum of its components is one. A matrix is a
stochastic matrix if each of its row vectors is stochastic. A stochastic
vector is positive if all its components are positive.

Let $B=(b_{ij})\in\mathbb{R}^{n\times n}$ be a stochastic matrix.
Define $\mathcal{G}_B=(\{1,...,n\},\mathcal{E}_B)$ as the
graph associated with $B$, where $(j,i)\in\mathcal{E}_B$ if and only
if $b_{ij}>0$ (its adjacency matrix is $B$). According to
Perron-Frobenius theorem \cite{hor}, there is a unique positive
stochastic left eigenvector of $B$ associated with eigenvalue one if
$\mathcal{G}_B$ is strongly connected. We call this eigenvector the
Perron vector of $B$.

\subsection{Convex Analysis}

A set $K\subseteq \mathbb{R}^m$ is convex if $\lambda z_1
+ (1-\lambda)z_2\in K$ for any $z_1, z_2 \in K$ and $0<\lambda<1$. A
point $z$ is an interior point of $K$ if
$\textbf{B}(z,\varepsilon)\subseteq K$ for some $\varepsilon>0$. For
a closed convex set $K$ in $\mathbb{R}^m$, we can associate with any
$z\in \mathbb{R}^m$ a unique element $P_K(z)\in K$ satisfying
$|z-P_K(z)|= \inf_{y\in K}|z-y|$, where $P_K$
is the projection operator onto $K$.
The following property for the convex projection operator $P_K$ holds by Lemma 1 (b) in \cite{nedic2},
\begin{align}\label{pro}
|P_K(y)-z|\leq|y-z|\;\mbox{ for any}\; y\in\mathbb{R}^m \;\mbox{and any}\; z\in K.
\end{align}

A function $\varphi(\cdot): \mathbb{R}^m\rightarrow \mathbb{R}$ is (strictly) convex if $
\varphi(\lambda z_1 + (1-\lambda)z_2)(<)\leq\lambda \varphi(z_1)
+ (1-\lambda)\varphi(z_2)$ for any $z_1\neq z_2 \in \mathbb{R}^m$ and $0<\lambda<1$.
A function $\varphi$ is (strictly) concave if $-\varphi$ is (strictly) convex.
A convex function $\varphi: \mathbb{R}^m\rightarrow \mathbb{R}$ is continuous.

For a convex function $\varphi$, $v(\hat z)\in\mathbb{R}^m$ is a
subgradient of $\varphi$ at point $\hat z$ if
$\varphi(z)\geq \varphi(\hat z)+\langle z-\hat z ,v(\hat z)\rangle$, $\forall z\in\mathbb{R}^m.$
For a concave function $\varphi$, $v(\hat
z)\in\mathbb{R}^m$ is a subgradient of $\varphi$ at $\hat z$ if
$\varphi(z)\leq \varphi(\hat z)+\langle z-\hat z ,v(\hat
z)\rangle,\forall z\in\mathbb{R}^m$. The set of all subgradients of (convex or
concave) function $\varphi$ at $\hat z$ is denoted by $\partial
\varphi(\hat z)$, which is called the
subdifferential of $\varphi$ at $\hat z$.

\subsection{Saddle Point and Nash Equilibrium}

A function $\phi(\cdot,\cdot):
\mathbb{R}^{m_1}\times\mathbb{R}^{m_2}\rightarrow \mathbb{R}$ is
(strictly) convex-concave if it is (strictly) convex in
first argument and (strictly) concave in second one. Given a
point $(\hat x, \hat y)$, we denote by $\partial_x \phi(\hat x,\hat y)$
the subdifferential of convex function $\phi(\cdot,\hat
y)$ at $\hat x$ and $\partial_y \phi(\hat x,\hat y)$ the
subdifferential of concave function $\phi(\hat x,\cdot)$ at $\hat y$.

A pair $(x^*,y^*)\in X\times Y$ is a saddle
point of $\phi$ on $X\times Y$ if
$$\phi(x^*,y)\leq \phi(x^*,y^*)\leq \phi(x,y^*), \forall x\in X, y\in Y.$$
The next lemma presents a necessary and sufficient condition to characterize the saddle points (see Proposition 2.6.1 in
\cite{nedbook}).

\begin{lem}
\label{lema1} Let $X\subseteq\mathbb{R}^{m_1}, Y\subseteq\mathbb{R}^{m_2}$ be two
closed convex sets. Then a pair $(x^*,y^*)$ is a saddle point of $\phi$ on $X\times Y$
if and only if
$$\sup_{y\in Y}\inf_{x\in X}\phi(x,y)=\inf_{x\in X}\sup_{y\in Y}\phi(x,y)=\phi(x^*,y^*),$$
and $x^*$ is an optimal solution of optimization problem
\begin{align}\label{pro1}
{\rm minimize}\;\; \sup_{y\in Y}\phi(x,y)\;\qquad {\rm subject\;to}\;\;x\in X,
\end{align}
while $y^*$ is an optimal solution of optimization problem
\begin{align}\label{pro2}
{\rm maximize}\;\; \inf_{x\in X}\phi(x,y)\;\qquad {\rm subject\;to}\;\;y\in Y.
\end{align}
\end{lem}

From Lemma \ref{lema1}, we find that all saddle points of $\phi$ on
$X\times Y$ yield the same value. The next lemma can be obtained from Lemma
\ref{lema1}.
\begin{lem}
\label{lem01} If $(x^*_1,y^*_1)$ and $(x^*_2,y^*_2)$ are two saddle
points of $\phi$ on $X\times Y$, then $(x^*_1,y^*_2)$ and $(x^*_2,y^*_1)$ are also saddle points
of $\phi$ on $X\times Y$.
\end{lem}

\begin{rem}
\label{lema2} Denote by $\bar Z$ the set of all saddle points of
function $\phi$ on $ X\times  Y$, $\bar X$ and $\bar Y$ the optimal solution sets of
optimization problems (\ref{pro1}) and (\ref{pro2}), respectively. Then
from Lemma \ref{lema1} it is not hard to find that if $\bar Z$ is nonempty, then
$\bar X$, $\bar Y$ are nonempty, convex, and $\bar Z=\bar X\times \bar Y$. Moreover, if
$X$ and $Y$ are convex, compact and $\phi$ is convex-concave, then $\bar Z$ is nonempty
(see Proposition 2.6.9 in
\cite{nedbook}).
\end{rem}

The saddle point computation can be related to a zero-sum game. In
fact, a (strategic) game is described as a triple
$(\mathcal{I},\mathcal{W},\mathcal{U}),$ where $\mathcal{I}$ is the
set of all players; $\mathcal{W}=\mathcal{W}_1\times\cdots \times
\mathcal{W}_n$, $n$ is the number of players, $\mathcal{W}_i$ is the
set of actions available to player $i$; $\mathcal{U}=(u_1,\dots,u_n)$,
$u_i:\mathcal{W}\rightarrow \mathbb{R}$ is the payoff function of
player $i$. The game is said to be zero-sum if $\sum^n_{i=1}u_i(w_i,w_{-i})=0$,
where $w_{-i}$ denotes the actions of all players other than $i$.
A profile action $w^*=(w^*_1,\dots,w^*_n)$ is said to be a Nash
equilibrium if $u_i(w^*_i,w^*_{-i})\geq u_i(w_i,w^*_{-i})$ for each
$i\in \mathcal{V}$ and $w_i \in \mathcal{W}_i$. The
Nash equilibria set of a two-person zero-sum game ($n=2, u_1+u_2=0$) is
exactly the saddle point set of payoff function $u_2$.

\section{Distributed Nash Equilibrium Computation}

In this section, we introduce a distributed Nash equilibrium
computation problem and then propose a subgradient-based algorithm as a solution.

Consider a network $\Xi$ consisting of two subnetworks $\Xi_1$ and
$\Xi_2$.   Agent $i$ in $\Xi_1$ is associated with a convex-concave
objective function $f_i(x,y):
\mathbb{R}^{m_1}\times\mathbb{R}^{m_2}\rightarrow \mathbb{R}$, and
agent $i$ in $\Xi_2$ is associated with a convex-concave
objective function $g_i(x,y):
\mathbb{R}^{m_1}\times\mathbb{R}^{m_2}\rightarrow \mathbb{R}$. Each
agent only knows its own objective function. The two subnetworks
have a common sum objective function with closed convex constraint
sets $X\subseteq \mathbb{R}^{m_1}, Y\subseteq \mathbb{R}^{m_2}$:
$$
U(x,y)=\sum^{n_1}_{i=1}f_i(x,y)=\sum^{n_2}_{i=1}g_i(x,y),\; x\in
X,\; y\in Y.
$$
Then the network is engaged in a (generalized) zero-sum game $\big(\{\Xi_1,\Xi_2\},X\times Y,u\big)$,
where $\Xi_1$ and $\Xi_2$
are viewed as two players, their respective payoff functions
are $u_{\Xi_1}=-\sum^{n_1}_{i=1}f_i$ and
$u_{\Xi_2}=\sum^{n_2}_{i=1}g_i$.
The objective of $\Xi_1$ and $\Xi_2$ is to achieve a
Nash equilibrium of the zero-sum game.

\begin{rem}
Despite that the contribution of this paper is mainly theoretical,
the considered model appears also in applications.
Here we illustrate that by discussing two practical examples in the literature.
In the first example, from \cite{cor2} note that
for multiple Gaussian communication channels with budget constrained signal power and noise levels,
the capacity of each channel is concave in signal power and convex in noise level.
Suppose there are two subnetworks, one of which is more critical than the other. The critical subnetwork aims to maximize its capacity by raising its transmission power while the other aims to reduce the interference to other channels by minimizing its transmission power
(and thus the capacity). The objective of the two subnetworks is then to find the Nash equilibrium of the sum of all channels' capacities, see Remark 3.1 in \cite{cor2} for more details.
For the second example, recall that many practical problems (for example, distributed estimation, resource allocation, optimal flow control) can be formulated as distributed convex constrained optimization problems, in which
the associated Lagrangian function can be expressed as a sum of individual Lagrangian functions,
which are convex in the optimization variable and linear (hence concave)
in the Lagrangian multiplier. Under Salter's condition, the optimal solutions
can be found by computing the saddle-points of the convex-concave Lagrangian function, or equivalently, the Nash equilibrium of the corresponding
zero-sum game, see \cite{zhu} for further discussions.
\end{rem}

We next provide a basic assumption.

\noindent {\bf A1} {\it(Existence of Saddle Points)}
For each stochastic vector $\mu$, $\sum^{n_1}_{i=1}\mu_if_i$ has at least one saddle point over $X\times Y$.

Clearly, {\bf A1} holds if $X$
and $Y$ are bounded (see Proposition 2.6.9 in
\cite{nedbook} for other conditions guaranteeing the existence of saddle points). However, in this paper we do not require
$X$ and $Y$ to be bounded. Let
$$Z^*=X^*\times
Y^*\subseteq X\times Y$$ denote the set of all saddle points of $U$ on $X\times Y$.
Notice that $X^*\times Y^*$ is
also the set of Nash equilibria of the generalized zero-sum game.

Denote the state of node $i\in \mathcal{V}_1$ as $x_i(k)\in \mathbb{R}^{m_1}$ and the state of node
$i\in \mathcal{V}_2$ as $y_i(k)\in \mathbb{R}^{m_2}$ at time $k=0,1,\dots$.

\begin{defn}
\label{def1} The network $\Xi$ is said to achieve a Nash equilibrium
 if, for any initial condition $x_i(0)\in
\mathbb{R}^{m_1},\,i\in \mathcal{V}_1$ and $y_i(0)\in
\mathbb{R}^{m_2},\,i\in \mathcal{V}_2$, there are $x^*\in X^*$ and
$y^*\in Y^*$ such that
$$
\lim_{k\rightarrow\infty}x_i(k)=x^*,\;i\in \mathcal{V}_1,\quad
\lim_{k\rightarrow\infty}y_i(k)=y^*,\;i\in \mathcal{V}_2.
$$
\end{defn}

The interconnection in the network $\Xi$ is time-varying and modeled
as three digraph sequences:
$$\mathcal{G}_1=\big\{\mathcal{G}_1(k)\big\},
\mathcal{G}_2=\big\{\mathcal{G}_2(k)\big\},
\mathcal{G}_{\bowtie}=\big\{\mathcal{G}_{\bowtie}(k)\big\},
$$
 where
$\mathcal{G}_1(k)=(\mathcal{V}_1,\mathcal{E}_1(k))$ and
$\mathcal{G}_2(k)=(\mathcal{V}_2,\mathcal{E}_2(k))$ are the graphs
to describe subnetworks $\Xi_1$ and $\Xi_2$, respectively, and
$\mathcal{G}_{\bowtie}(k)=(\mathcal{V},\mathcal{E}_{\bowtie}(k))$ is
the bipartite graph to describe the interconnection between $\Xi_1$
and $\Xi_2$ at time $k\geq0$. For $k_2>k_1\geq0$, denote
by $\mathcal{G}_{\bowtie}\big([k_1,k_2)\big)$ the union
graph with node set $\mathcal{V}$ and arc set
$\bigcup^{k_2-1}_{s=k_1}\mathcal{E}_{\bowtie}(s)$, and
$\mathcal{G}_\ell\big([k_1,k_2)\big)$ the union graph
with node set $\mathcal{V}_\ell$ and arc set
$\bigcup^{k_2-1}_{s=k_1}\mathcal{E}_\ell(s)$ for $\ell=1,2$.
The following assumption on
connectivity is made.

\noindent {\bf A2} {\it (Connectivity)} (i) The graph sequence
$\mathcal{G}_{\bowtie}$ is uniformly jointly bipartite; namely, there is an integer $T_{\bowtie}>0$ such that
$\mathcal{G}_{\bowtie}\big([k,k+T_{\bowtie})\big)$ is bipartite without isolated nodes for $k\geq 0$.

(ii) For $\ell=1,2$, the graph sequence $\mathcal{G}_\ell$ is
uniformly jointly strongly connected (UJSC); namely, there
is an integer $T_\ell>0$ such that
$\mathcal{G}_\ell\big([k,k+T_\ell)\big)$ is strongly connected
for $k\geq 0$.

\begin{rem}
The agents in
$\Xi_\ell$ connect directly with those in $\Xi_{3-\ell}$ for all
the time in \cite{cor2}, while the agents in two subnetworks are
connected at least once in each interval of length $T_{\bowtie}$
according to {\bf {A2}} $(i)$. In fact, it may be practically hard
for the agents of different subnetworks to maintain communications all
the time. Moreover, even if each agent in $\Xi_\ell$ can receive the
information from $\Xi_{3-\ell}$, agents may just send or receive once during a period of length $T_{\bowtie}$ to save energy or communication cost.
\end{rem}

To handle the distributed Nash equilibrium computation problem, we
propose a subgradient-based algorithm, called
\emph{Distributed Nash Equilibrium Computation Algorithm}:
\begin{equation}
\label{6} \begin{cases} &x_i(k+1)=P_X\big(\hat x_i(k)-\alpha_{i,k}q_{1i}(k)\big),\\
&\qquad\qquad q_{1i}(k)\in\partial_xf_i\big(\hat x_i(k),\breve{x}_i(k)\big),\;i\in\mathcal{V}_1,\\
&y_i(k+1)=P_Y\big(\hat y_i(k)+\beta_{i,k}q_{2i}(k)\big),\\
&\qquad\qquad  q_{2i}(k)\in\partial_yg_i\big(\breve{y}_i(k),
 \hat y_i(k)\big),\;i\in\mathcal{V}_2\end{cases}
\end{equation}
with
\begin{align}
&\hat x_i(k)=\sum_{j\in\mathcal{N}^1_i(k)}a_{ij}(k)x_j(k),\;
\breve{x}_i(k)=\sum_{j\in\mathcal{N}^{2}_i(\breve{k}_i)}a_{ij}(\breve{k}_i)y_j(\breve{k}_i),\nonumber\\
&\hat y_i(k)=\sum_{j\in\mathcal{N}^2_i(k)}a_{ij}(k)y_j(k),\;
\breve{y}_i(k)=\sum_{j\in\mathcal{N}^{1}_i(\breve{k}_i)}a_{ij}(\breve{k}_i)x_j(\breve{k}_i),\nonumber
\end{align}
where
$\alpha_{i,k}>0$, $\beta_{i,k}>0$ are the stepsizes at time $k$,
$a_{ij}(k)$ is the time-varying weight of arc $(j,i)$,
$\mathcal{N}^\ell_i(k)$ is the set of neighbors in $\mathcal{V}_\ell$
of node $i$ at time $k$,
and
\begin{align}
\label{8} \breve{k}_i=\max\big\{s|s\leq k,
\mathcal{N}^{3-\ell}_i(s)\neq\emptyset\big\}\leq k,
\end{align}
which is the
last time before $k$ when node $i\in \mathcal{V}_\ell$ has at least
one neighbor in $\mathcal{V}_{3-\ell}$.

\begin{figure}[!htbp]
\centering
\includegraphics[width=2.4in]{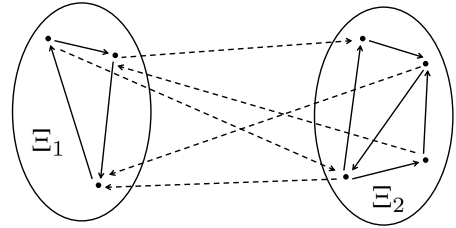}
\caption{The zero-sum game communication graph}
\end{figure}

\begin{rem}
When all objective functions $f_i,g_i$ are additively
separable, i.e., $f_i(x,y)=f^1_i(x)+f^2_i(y),\;g_i(x,y)=g^1_i(x)+g^2_i(y)$,
the considered distributed Nash equilibrium computation problem is equivalent to two
separated distributed optimization problems with respective objective functions $\sum^{n_1}_{i=1}f^1_i(x)$,
$\sum^{n_2}_{i=1}g^2_i(y)$ and constraint sets $X$, $Y$. In this case, the set of Nash equilibria is given by
$$
X^*\times Y^*=\arg\min_X\sum^{n_1}_{i=1}f^1_i\times\arg\max_Y\sum^{n_2}_{i=1}g^2_i.
$$
Since $\partial_xf_i(x,y)=\partial_xf^1_i(x)$ and
$\partial_yg_i(x,y)=\partial_yg^2_i(y)$, algorithm (\ref{6}) becomes in this case the well-known distributed subgradient algorithms \cite{nedic1,nedic2}.
\end{rem}

\begin{rem}
To deal with \emph{weight-unbalanced} graphs,
some methods, the rescaling technique \cite{ols} and the push-sum protocols \cite{kem,ben,tsi1} have been proposed for average consensus problems; reweighting the objectives \cite{tsi3} and the subgradient-push protocols \cite{ned,nedic5}
for distributed optimization problems. Different from these methods, in this paper we propose a
distributed algorithm to handle weight-unbalanced graphs
when the stepsizes taken by agents are not necessarily the same.
\end{rem}

\begin{rem}
Different from the extreme-seeking techniques used in \cite{fri,sta},
our method uses the subgradient to compute the Nash equilibrium.
\end{rem}

The next assumption was
also used in \cite{nedic1,nedic2,zhu,shi1}.

\noindent{\bf A3} {\it (Weight Rule)} (i) There is $0<\eta<1$ such
that $a_{ij}(k)\geq \eta$ for all $i, k$ and
$j\in\mathcal{N}^1_i(k)\bigcup\mathcal{N}^2_i(k)$;

(ii) $\sum_{j\in\mathcal{N}^{\ell}_i(k)}a_{ij}(k)=1$
for all $k$ and $i\in \mathcal{V}_\ell,\ell=1,2$;

(iii)
$\sum_{j\in\mathcal{N}^{3-\ell}_i(\breve{k}_i)}a_{ij}(\breve{k}_i)=1$
for $i\in \mathcal{V}_\ell,\ell=1,2.$

Conditions (ii) and (iii) in {\bf A3} state that the information from an agent's neighbors is
used through a weighted average.
The next assumption is about subgradients of objective functions.

\noindent {\bf A4} {\it (Boundedness of Subgradients)} There is
$L>0$ such that, for each $i,j$,
$$
|q|\leq L, \;\; \forall
q\in\partial_xf_i(x,y)\bigcup\partial_yg_j(x,y), \;\forall x\in X, y\in
Y.$$

Obviously, {\bf A4} holds
if $X$ and $Y$ are bounded. A similar bounded assumption has been widely used in distributed optimization
\cite{joh09,nedic,nedic1,nedic2}.

Note that the stepsize in our algorithm (\ref{6})
is {\it heterogenous}, i.e., the stepsizes may be different for
different agents, in order to deal with general unbalanced cases.
One challenging  problem is how to select the stepsizes
$\{\alpha_{i,k}\}$ and $\{\beta_{i,k}\}$.  The
{\it homogenous} stepsize case is to set
$\alpha_{i,k}=\beta_{j,k}=\gamma_k$ for $i\in \mathcal{V}_1,j\in
\mathcal{V}_2$ and all $k$, where
$\{\gamma_k\}$ is given as follows.

\noindent {\bf A5} $\{\gamma_k\}$ is non-increasing,
$\sum^{\infty}_{k=0}\gamma_k=\infty$ and
$\sum^{\infty}_{k=0}\gamma^2_k<\infty$.

Conditions $\sum^{\infty}_{k=0}\gamma_k=\infty$ and
$\sum^{\infty}_{k=0}\gamma^2_k<\infty$ in {\bf A5} are
well-known in homogeneous stepsize selection for distributed
subgradient algorithms
for distributed optimization problems with weight-balanced graphs, e.g., \cite{nedic2,ram,zhu}.

\begin{rem}
While weight-balanced graphs are considered in \cite{nedic1,nedic2,zhu,cor2,cor14},
we consider general (weight-unbalanced) digraphs, and provide a heterogeneous
stepsize design method for the desired Nash equilibrium convergence.
\end{rem}

\section{Main Results}

In this section, we start with homogeneous stepsizes to achieve
a Nash equilibrium for weight-balanced graphs (in Section IV.A). Then
we focus on a special weight-unbalanced case to show how a homogeneous-stepsize
algorithm may fail to achieve our aim (in
Section IV.B). Finally, we show that the heterogeneity of
stepsizes can help us achieve a Nash equilibrium in some
weight-unbalanced graph cases (in Section IV.C).

\subsection{Weight-balanced Graphs}

Here we consider algorithm (\ref{6}) with
homogeneous stepsizes $\alpha_{i,k}=\beta_{i,k}=\gamma_k$ for weight-balanced digraphs.
The following result, in fact, provides two sufficient conditions to
achieve a Nash equilibrium under switching weight-balanced digraphs.

\begin{them}
\label{thm1} Suppose {\bf A1}--{\bf A5} hold and digraph
$\mathcal{G}_\ell(k)$ is weight-balanced for $k\geq 0$ and
$\ell=1,2$. Then the multi-agent network $\Xi$ achieves a Nash
equilibrium by algorithm (\ref{6}) with the
homogeneous stepsizes $\{\gamma_k\}$ if either of the following
two conditions holds:

(i)  $U$ is strictly convex-concave;

(ii) $X^*\times Y^*$ contains an interior point.
\end{them}

The proof can be found in Section V.B.

\begin{rem}
The authors in \cite{cor2} developed a continuous-time dynamical
system to solve the Nash equilibrium computation problem for
fixed weight-balanced digraphs, and showed that the network
converges to a Nash equilibrium for a strictly
convex-concave differentiable sum objective function.
Different from \cite{cor2}, here we allow time-varying communication structures and a non-smooth  objective function $U$. The same result may also hold for the continuous-time solution in \cite{cor2} under our problem setup, but the analysis would probably be much more involved.
\end{rem}

\subsection{Homogenous Stepsizes vs. Unbalanced Graphs}

In the preceding subsection, we showed that a Nash equilibrium can be achieved
with homogeneous stepsizes when the graphs of two
subnetworks are weight-balanced.  Here we demonstrate that the homogenous stepsize algorithm may fail
to guarantee the Nash equilibrium convergence for general
weight-unbalanced digraphs unless certain conditions about the
objective function hold.

Consider a special case, called the completely identical subnetwork
case, i.e., $\Xi_1$ and $\Xi_2$ are completely identical:
\begin{align}
&n_1=n_2,\;f_i=g_i,\;i=1,...,n_1;\;A_1(k)=A_2(k),\nonumber\\
&\mathcal{G}_{\bowtie}(k)=\big\{(\omega^\ell_i,\omega^{3-\ell}_i),\ell=1,2,i=1,...,n_1\big\},k\geq 0.\nonumber
\end{align}
In this case, agents $\omega^\ell_i,\omega^{3-\ell}_i$
have the same objective function, neighbor set and can communicate with each other at all times.
Each pair of agents $\omega^\ell_i,\omega^{3-\ell}_i$ can be viewed as one agent labeled as ``$i$''.
Then algorithm (\ref{6}) with homogeneous stepsizes $\{\gamma_k\}$ reduces to
the following form:
\begin{align}\label{spe}
\begin{cases}x_i(k+1)=P_X\big(\sum_{j\in\mathcal{N}^1_i(k)}a_{ij}(k)x_j(k)-\gamma_kq_{1i}(k)\big), \\
y_i(k+1)=P_Y\big(\sum_{j\in\mathcal{N}^1_i(k)}a_{ij}(k)y_j(k)+\gamma_kq_{2i}(k)\big), \\
 \end{cases}
\end{align}
for $i=1,...,n_1$, where 
$q_{1i}(k)\in\partial_xf_i(\hat x_i(k),y_i(k))$,
$q_{2i}(k)\in\partial_yf_i(x_i(k),\hat y_i(k))$.

\begin{rem}
Similar distributed saddle point computation algorithms have been
proposed in the literature, for example, the distributed saddle point computation
for the Lagrange function of constrained optimization problems in
\cite{zhu}.
In fact, algorithm (\ref{spe}) can be used to solve the following distributed saddle-point computation problem: consider a network $\Xi_1$ consisting of $n_1$ agents with node
set $\mathcal{V}_1=\{1,...,n_1\}$, its objective is
to seek a saddle point of the sum objective function
$\sum^{n_1}_{i=1}f_i(x,y)$ in a distributed way, where $f_i$ can only be known by agent
$i$. In (\ref{spe}), $(x_i,y_i)$ is the state of node ``$i$''.
Moreover,  algorithm (\ref{spe}) can be viewed as a
distributed version of the following centralized algorithm:
\begin{align}
\begin{cases}x(k+1)=P_X\big( x(k)-\gamma q_1(k)\big),q_1(k)\in \partial_x U(x(k), y(k)),\nonumber\\
y(k+1)=P_Y\big(y(k)+\gamma q_2(k)\big),q_2(k)\in \partial_y
U(x(k),y(k)),\nonumber
\end{cases}
\end{align}
which was proposed in \cite{nedic} to solve the approximate saddle point problem with a constant stepsize.
\end{rem}

We first show that, algorithm (\ref{6}) with homogeneous
stepsizes (or equivalently (\ref{spe})) cannot seek the desired Nash
equilibrium though it is convergent, even for {\em fixed}
weight-unbalanced graphs.

\begin{them}\label{th6.3}
Suppose {\bf A1}, {\bf A3}--{\bf A5} hold, and $f_i,i=1,...,n_1$ are
strictly convex-concave and the graph is fixed with
$\mathcal{G}_1(0)$ strongly connected.
Then, with
(\ref{spe}), all the agents converge to the unique saddle point,
denoted as $(\vec{x}, \vec{y})$, of an objective function
$\sum^{n_1}_{i=1}\mu_if_i$ on $X\times Y$, where
$\mu=(\mu_1,\dots,\mu_{n_1})'$ is the Perron vector of the adjacency
matrix $A_1(0)$ of graph $\mathcal{G}_1(0)$.
\end{them}

The proof is almost the same as
that of Theorem \ref{thm1}, by replacing $\sum^{n_1}_{i=1}|x_i(k)-x^*|^2,\sum^{n_2}_{i=1}|y_i(k)-y^*|^2$ and
$U(x,y)$ with
$\sum^{n_1}_{i=1}\mu_i|x_i(k)-\vec{x}|^2,\sum^{n_1}_{i=1}\mu_i|y_i(k)-\vec{y}|^2$
and $\sum^{n_1}_{i=1}\mu_if_i(x,y)$, respectively.  Therefore, the proof is omitted.

Although it is hard to achieve the desired Nash equilibrium with the
homogeneous-stepsize algorithm in general, we can still achieve it in some
cases. Here we can give a necessary and sufficient condition to
achieve a Nash
equilibrium for any UJSC switching digraph sequence. 

\begin{them}
\label{thmdao1}
Suppose {\bf A1}, {\bf A3}--{\bf A5} hold and $f_i,i=1,...,n_1$ are strictly
convex-concave. Then the multi-agent network $\Xi$ achieves a Nash
equilibrium by algorithm (\ref{spe}) for any UJSC switching digraph sequence $\mathcal{G}_1$
 if and only if
$f_i,i=1,...,n_1$ have the same saddle point on $X\times Y$.
\end{them}

The proof can be found in Section V.C.

\begin{rem}
The strict convexity-concavity of $f_i$ implies that the saddle point of $f_i$ is unique.
From the proof we can find that the necessity of Theorem  \ref{thmdao1} does
not require that each objective function $f_i$ is
strictly convex-concave, but the strict convexity-concavity of the sum objective function $\sum^{n_1}_{i=1}f_i$
suffices.
\end{rem}

\subsection{Weight-unbalanced Graphs}

The results in the preceding subsections showed that the homogenous-stepsize algorithm may not make a weight-unbalanced network achieve its Nash equilibrium.
Here we first show the existence of a heterogeneous-stepsize design to make the (possibly weight-unbalanced) network
achieve a Nash equilibrium.

\begin{them}
\label{th4.4} Suppose {\bf A1}, {\bf A3}, {\bf A4} hold and $U$ is strictly
convex-concave. Then for any time-varying communication graphs
$\mathcal{G}_\ell,\ell=1,2$ and $\mathcal{G}_{\bowtie}$ that satisfy {\bf A2}, there always exist stepsize sequences
$\{\alpha_{i,k}\}$ and $\{\beta_{i,k}\}$ such that the multi-agent
network $\Xi$ achieves a Nash equilibrium by algorithm (\ref{6}).
\end{them}

The proof is in Section V.D.
In fact, it suffices to design stepsizes $\alpha_{i,k}$ and $\beta_{i,k}$
as follows:
\begin{align}\label{desi2}
\alpha_{i,k}=\frac{1}{\alpha^i_k}\gamma_k,\;\;
\beta_{i,k}=\frac{1}{\beta^i_k}\gamma_k,
\end{align}
 where $(\alpha^1_k,\dots,\alpha^{n_1}_k)'=\phi^1(k+1)$, $(\beta^1_k,\dots,\beta^{n_2}_k)'=\phi^2(k+1)$,
 $\phi^\ell(k+1)$ is the vector for which
$\lim_{r\rightarrow\infty}\Phi^\ell(r,k+1):=\textbf{1}(\phi^\ell(k+1))'$,
$\Phi^\ell(r,k+1):=A_\ell(r)A_\ell(r-1)\cdots A_\ell(k+1)$,
$\ell=1,2$,
 $\{\gamma_k\}$ satisfies
the following conditions:
\begin{equation}\label{desi1}
\begin{aligned}
&\lim_{k\rightarrow\infty}\gamma_k\sum^{k-1}_{s=0}\gamma_s=0,\;
\{\gamma_k\}\;\mbox{is non-increasing},\\
&\sum^\infty_{k=0}\gamma_k=\infty,\;\sum^\infty_{k=0}\gamma^2_k<\infty.
\end{aligned}
\end{equation}

\begin{rem}\label{rem4.4}
The stepsize design in Theorem \ref{th4.4} is motivated by the following
two ideas. On one hand, agents need to eliminate the imbalance caused by the weight-unbalanced
graphs, which is done by $\{1/\alpha^i_k\},\{1/\beta^i_k\}$,
while on the other hand, agents also need to achieve a consensus within each subnetwork
and cooperative optimization, which is done by $\{\gamma_k\}$, as in the balanced graph case.
\end{rem}

\begin{rem}
Condition (\ref{desi1}) can be satisfied by letting
$\gamma_k=\frac{c}{(k+b)^{\frac{1}{2}+\epsilon}}$ for $k\geq0$, $c>0$, $b>0$,
$0<\epsilon\leq \frac{1}{2}$. Moreover, from the proof of
Theorem \ref{th4.4} we find that, if the sets $X$ and $Y$ are
bounded, the system states are naturally
bounded, and then (\ref{desi1}) can be relaxed as {\bf A5}.
\end{rem}

Clearly, the above choice of stepsizes at time $k$ depend
on the adjacency matrix sequences $\{A_1(s)\}_{s\geq k+1}$ and
$\{A_2(s)\}_{s\geq k+1}$, which is not so practical.  Therefore,
we will consider how to design adaptive algorithms to update the stepsize sequences
$\{\alpha_{i,k}\}$ and $\{\beta_{i,k}\}$ such that the Nash
equilibrium can be achieved, where the (heterogeneous) stepsizes at time $k$ just
depend on the local information that agents can obtain before time $k$.

Take
\begin{align}\label{desi3}
\alpha_{i,k}=\frac{1}{\hat \alpha^i_k}\gamma_k,
\;\;\beta_{i,k}=\frac{1}{\hat \beta^i_k}\gamma_k,
\end{align}
where $\{\gamma_k\}$ satisfies (\ref{desi1}). The only difference between
stepsize   selection rule (\ref{desi3}) and (\ref{desi2})
is that $\alpha^i_k$ and $\beta^i_k$ are replaced with $\hat \alpha^i_k$ and $\hat \beta^i_k$, respectively.
We consider how
to design distributed adaptive algorithms for $\hat \alpha^i$ and
$\hat \beta^i$ such that
\begin{equation}\label{conn}
\begin{aligned}
&\hat\alpha^i_k=\hat\alpha^i\big(a_{ij}(s),j\in \mathcal{N}^1_i(s),s\leq k\big),\\
&\hat\beta^i_k=\hat\beta^i\big(a_{ij}(s),j\in \mathcal{N}^2_i(s),s\leq k\big),
\end{aligned}
\end{equation}
and
\begin{align}\label{co1}
\lim_{k\rightarrow\infty}\big(\hat \alpha^i_k-\alpha^i_k\big)=0,\;
\lim_{k\rightarrow\infty}\big(\hat \beta^i_k-\beta^i_k\big)=0.
\end{align}
Note that $(\alpha^1_k,\dots,\alpha^{n_1}_k)'$ and $(\beta^1_k,\dots,\beta^{n_2}_k)'$
are the Perron vectors of the two limits $\lim_{r\rightarrow\infty}\Phi^1(r,k+1)$
and $\lim_{r\rightarrow\infty}\Phi^2(r,k+1)$, respectively.

The next theorem shows that, in two standard cases, we can design
distributed adaptive algorithms satisfying (\ref{conn}) and
(\ref{co1}) to ensure that $\Xi$ achieves a Nash
equilibrium.  How to design them is given in the proof.

\begin{them}
\label{th4.5} Consider algorithm (\ref{6}) with stepsize
selection rule (\ref{desi3}). Suppose {\bf A1}--{\bf A4} hold, $U$
is strictly convex-concave. For the following two cases, with the
adaptive distributed algorithms satisfying (\ref{conn}) and
(\ref{co1}), network $\Xi$ achieves a Nash equilibrium.

(i) For $\ell=1,2$, the adjacency matrices $A_\ell(k),k\geq0$ have a common left eigenvector with eigenvalue one;

(ii) For $\ell=1,2$, the adjacency matrices $A_\ell(k),k\geq 0$ are switching periodically, i.e., there exist
positive integers $p^\ell$ and two finite sets of stochastic matrices $A^0_\ell,...,A^{p^\ell-1}_\ell$ such that
$A_\ell(rp^\ell+s)=A^s_\ell$ for $r\geq 0$ and $s=0,...,p^\ell-1$.
\end{them}

The proof is given in Section V.E.

\begin{rem}
Regarding case (i), note that for a {\em fixed} graph, the adjacency matrices obviously have a common left eigenvector.
Moreover, periodic switching can be interpreted as a simple scheduling strategy.
At each time agents may choose some neighbors to communicate with in a periodic order.
\end{rem}

\begin{rem}
In the case of a fixed unbalanced graph,
the optimization can also be
solved by either reweighting the objectives \cite{tsi3}, or by the subgradient-push protocols \cite{ned,nedic5}, where the Perron vector of the adjacency
matrix is required to be known in advance or each agent is required to know its out-degree.
These requirements may be quite restrictive in a distributed setting.
Theorem \ref{th4.5} shows that, in the fixed graph case, agents can
adaptively learn the Perron vector by the adaptive learning scheme and then achieve the desired convergence
without knowing the Perron vector and their individual out-degrees.
\end{rem}

When the adjacency matrices $A_\ell(k)$ have a common left eigenvector,
the designed distributed adaptive learning strategy (\ref{de}) can guarantee that the differences between
$\hat \alpha^i_k=\alpha^i_i(k)$, $\hat \beta^i_k=\beta^i_i(k)$ and the
``true stepsizes'' $\phi^1_i(k+1)$, $\phi^2_i(k+1)$ asymptotically tend to zero.
The converse is also true for some cases.
In fact, if the time-varying adjacency matrices are switching within finite matrices
and $\lim_{k\rightarrow\infty}(\alpha^i_i(k)-\phi^1_i(k+1))=0$, $\lim_{k\rightarrow\infty}(\beta^i_i(k)-\phi^2_i(k+1))=0$, then we can show that
the finite adjacency matrices certainly have a common left eigenvector.

Moreover, when the adjacency matrices have no common left eigenvector, the adaptive
learning strategy (\ref{de}) generally cannot make $\hat \alpha^i_k$, $\hat \beta^i_k$
asymptotically learn the true stepsizes and then cannot achieve a Nash equilibrium.
For instance, consider the special distributed saddle-point computation algorithm (\ref{spe})
with strictly convex-concave objective functions $f_i$.
Let $\bar\alpha=(\bar\alpha_1,...,\bar\alpha_{n_1})',\hat\alpha=(\hat\alpha_1,...,\hat\alpha_{n_1})'$ be two different positive stochastic vectors. Suppose $A_1(0)=\textbf{1}\bar\alpha'$ and
$A_1(k)=\textbf{1}\hat \alpha'$ for $k\geq1$. In this case, $\alpha^i_i(k)=\bar \alpha_i$,
$\phi^1_i(k+1)=\hat \alpha_i$ for all $k\geq0$ and then (\ref{co1}) is not true.
According to Theorem \ref{th6.3}, the learning strategy (\ref{de})
can make $(x_i(k),y_i(k))$ converge to the (unique) saddle point of
the function $\sum^{n_1}_{i=1}\frac{\hat \alpha_i}{\bar \alpha_i}f_i(x,y)$ on $X\times Y$, which is not necessarily
the saddle point of $\sum^{n_1}_{i=1}f_i(x,y)$ on $X\times Y$.

\section{Proofs}
In this section, we first introduce some useful lemmas and then
present the proofs of the theorems in last section.

\subsection{Supporting Lemmas}

First of all, we introduce two lemmas. The
first lemma is the deterministic version of Lemma 11 on page 50 in
\cite{Plo}, while the second one is Lemma 7 in \cite{nedic2}.

\begin{lem}
\label{4.1} Let $\{a_k\}$, $\{b_k\}$ and $\{c_k\}$ be non-negative
sequences with $\sum^{\infty}_{k=0}b_k<\infty$. If $ a_{k+1}\leq
a_k+b_k-c_k$ holds for any $k$, then $\lim_{k\rightarrow\infty}a_k$
is a finite number.
\end{lem}

\begin{lem}
\label{4.2} Let $0<\lambda<1$ and $\{a_k\}$ be a positive sequence.
If $\lim_{k\rightarrow\infty}a_k=0$,
then $\lim_{k\rightarrow\infty}\sum^k_{r=0}\lambda^{k-r}a_r=0$.
Moreover, if $\sum^\infty_{k=0}a_k<\infty$, then
$\sum^\infty_{k=0}\sum^k_{r=0}\lambda^{k-r}a_r<\infty$.
\end{lem}

Next, we show some useful lemmas.

\begin{lem}
\label{4.3} For any
$\mu\in\mathcal{S}^+_{n}$, there is a
stochastic matrix $B=(b_{ij})\in\mathbb{R}^{n\times n}$ such that $\mathcal{G}_B$
is strongly connected and $\mu' B=\mu'$.
\end{lem}
\emph{Proof:} Take $\mu=(\mu_1,\dots,\mu_n)'\in\mathcal{S}^+_{n}$. Without
loss of generality, we assume $\mu_1=\min_{1\leq i\leq n}\mu_i$
(otherwise we can rearrange the index of agents). Let $B$ be a
stochastic matrix such that the graph $\mathcal{G}_B$ associated
with $B$ is a directed cycle: $1e_nn\cdots 2e_11$ with $e_r=(r+1,
r), 1\leq r\leq n-1$ and $e_n=(1,n)$. Clearly, $\mathcal{G}_B$ is
strongly connected. Then all nonzero entries of $B$ are
$\big\{b_{ii},b_{i(i+1)},1\leq i\leq n-1,b_{nn},b_{n1}\big\}$ and
$\mu' B=\mu'$ can be rewritten as
$b_{11}\mu_1+(1-b_{nn})\mu_n=\mu_1$,
$(1-b_{rr})\mu_{r}+b_{(r+1)(r+1)}\mu_{r+1}=\mu_{r+1},1\leq
r\leq n-1$. Equivalently,
\begin{equation}\label{ss}
\left\{
  \begin{array}{ll}
    \;(1-b_{22})\mu_{2}=(1-b_{11})\mu_{1}\\
    \;(1-b_{33})\mu_{3}=(1-b_{11})\mu_{1} \\
    \qquad\qquad\;\;\;\;\;\;\vdots \\
    (1-b_{nn})\mu_{n}=(1-b_{11})\mu_1
  \end{array}
\right.
\end{equation}
Let $b_{11}=b^*_{11}$ with $0<b^*_{11}<1$. Clearly, there is a
solution to (\ref{ss}):
$b_{11}=b^*_{11},0<b_{rr}=1-(1-b^*_{11})\mu_1/\mu_r<1,2\leq r\leq
n$. Then the conclusion follows. \hfill$\square$

The following lemma is about stochastic matrices,
which can be found from Lemma 3 in \cite{chen}.

\begin{lem}\label{cons}
Let $B=(b_{ij})\in \mathbb{R}^{n\times n}$ be a stochastic matrix
and $\hbar(\mu)=\max_{1\leq i,j\leq n}|\mu_i-\mu_j|,
\mu=(\mu_1,\dots,\mu_n)'\in \mathbb{R}^n.$ Then
$\hbar(B\mu)\leq\mu(B)\hbar(\mu),$ where
$\mu(B)=1-\min_{j_1,j_2}\sum^n_{i=1}\min\{b_{j_1i},b_{j_2i}\}$, is
called ``the ergodicity coefficient" of $B$.
\end{lem}

We next give a lemma about the transition matrix sequence
$\Phi^\ell(k,s)=A_\ell(k)A_\ell(k-1)\cdots A_\ell(s), k\geq s$,
$\ell=1,2$, where (i), (ii) and (iv) are taken from Lemma 4 in
\cite{nedic1}, while (iii) can be obtained from Lemma 2 in
\cite{nedic1}.

\begin{lem}\label{lem1}
Suppose {\bf A2} (ii) and {\bf A3} (i), (ii) hold. Then for $\ell=1,2$, we have

(i) The limit $\lim_{k\rightarrow\infty}\Phi^\ell(k,s)$ exists for
each $s$;

(ii)
There is a positive stochastic vector $\phi^\ell(s)=(\phi_1^\ell(s),...,\phi_{n_\ell}^\ell(s))'$ such that
$\lim_{k\rightarrow\infty}\Phi^\ell(k,s)=\textbf{1}(\phi^\ell(s))'$;

(iii) For every $i=1,...,n_\ell$ and $s$, $\phi^\ell_i(s)\geq \eta^{(n_\ell-1)T_\ell}$;

(iv) For every $i$, the entries $\Phi^\ell(k,s)_{ij},j=1,...,n_\ell$ converge to the same limit
$\phi^\ell_j(s)$ at a geometric rate, i.e., for every $i=1,...,n_\ell$ and all $s\geq 0$,
$$
\big|\Phi^\ell(k,s)_{ij}-\phi^\ell_j(s)\big|\leq C_\ell \rho^{k-s}_\ell
$$
for all $k\geq s$ and $j=1,...,n_\ell$, where
$C_\ell=2\frac{1+\eta^{-M_\ell}}{1-\eta^{M_\ell}}$,
$\rho_\ell=(1-\eta^{M_\ell})^{\frac{1}{M_\ell}}$, and
$M_\ell=(n_\ell-1)T_\ell$.
\end{lem}

The following lemma shows a relation between the left
eigenvectors of stochastic matrices and the Perron vector of the
limit of their product matrix.

\begin{lem}\label{l1}
Let $\{B(k)\}$ be a sequence of stochastic matrices. Suppose
$B(k),k\geq 0$ have a common left eigenvector $\mu$ corresponding to eigenvalue
one and the associated graph sequence $\{\mathcal{G}_{B(k)}\}$ is
UJSC. Then, for each $s$,
$$
\lim_{k\rightarrow\infty}B(k)\cdots
B(s)=\textbf{1}\mu'/(\mu'\textbf{1}).
$$
\end{lem}

\emph{Proof}: Since $\mu$ is the common left eigenvector of
$B(r),r\geq s$ associated with eigenvalue one,
$\mu'\lim_{k\rightarrow\infty}B(k)\cdots
B(s)=\lim_{k\rightarrow\infty}\mu' B(k)\cdots B(s)=\mu'$. In addition,
by Lemma \ref{lem1}, for each $s$, the limit $\lim_{k\rightarrow\infty}B(k)\cdots B(s):=\textbf{1}\phi'(s)$
exists. Therefore,
$\mu'=\mu'(\textbf{1}\phi'(s))=(\mu'\textbf{1})\phi'(s)$, which implies $(\mu'\textbf{1})\phi(s)=\mu$.
The conclusion follows. \hfill$\square$

Basically, the two dynamics of algorithm (\ref{6}) are in the same
form.  Let us check the first one,
\begin{align}
\label{xi-eq} x_i(k+1)&=P_X\big(\hat
x_i(k)-\alpha_{i,k}q_{1i}(k)\big),\nonumber\\
&\quad q_{1i}(k)\in\partial_xf_i\big(\hat
x_i(k),\breve{x}_i(k)\big),\;i\in\mathcal{V}_1.
\end{align}
By treating the term containing $y_j \; (j\in\mathcal{V}_2)$ as
``disturbance", we can transform (\ref{xi-eq}) to a simplified model
in the following form with disturbance $\epsilon_i$:
\begin{align}\label{0con}
x_i(k+1)=\sum_{j\in\mathcal{N}^1_i(k)}a_{ij}(k)x_j(k)+\epsilon_i(k),\;i\in\mathcal{V}_1,
\end{align}
where $\epsilon_i(k)=P_X\big(\hat x_i(k)+w_i(k)\big)-\hat x_i(k)$. It follows from $x_j(k)\in X$,
the convexity of $X$ and {\bf A3} (ii) that $\hat x_i(k)=\sum_{j\in\mathcal{N}^1_i(k)}a_{ij}(k)x_j(k)\in
X$.
Then from (\ref{pro}), $|\epsilon_i(k)|\leq |w_i(k)|$.

The next lemma is about a limit for the two subnetworks.
Denote
$$
\bar \alpha_k=\max_{1\leq i\leq n_1}\alpha_{i,k},\;\;\bar \beta_k=\max_{1\leq i\leq n_2}\beta_{i,k}.
$$
\begin{lem}\label{le2}
Consider algorithm (\ref{6}) with {\bf A3} (ii) and {\bf A4}. If
$\lim_{k\rightarrow\infty}\bar\alpha_k\sum^{k-1}_{s=0}\bar\alpha_s
=\lim_{k\rightarrow\infty}$ $\bar\beta_k\sum^{k-1}_{s=0}\bar\beta_s=0$, then for any $x,y$,
$\lim_{k\rightarrow\infty}\bar \alpha_k\max_{1\leq i\leq n_1}|x_i(k)-x|=
\lim_{k\rightarrow\infty}\bar \beta_k\max_{1\leq i\leq n_2}|y_i(k)-y|=0.$
\end{lem}

\emph{Proof}:
We will only show $\lim_{k\rightarrow\infty}\bar \alpha_k\max_{1\leq i\leq n_1}|x_i(k)-x|=0$
 since the other one about $\bar \beta_k$ can be proved similarly.
At first, it follows from $\lim_{k\rightarrow\infty}\bar
\alpha_k\sum^{k-1}_{s=0}\bar \alpha_s=0$ that
$\lim_{k\rightarrow\infty}\bar \alpha_k=0$.
From {\bf A4} we have $|\epsilon_i(k)|\leq \bar \alpha_kL$.
Then from (\ref{0con}) and {\bf A3} (ii) we obtain
$$\max_{1\leq i\leq n_1}|x_i(k+1)-x|\leq\max_{1\leq i\leq n_1}|x_i(k)-x|+\bar
\alpha_kL,\forall k.$$
 Therefore, $\max_{1\leq i\leq
n_1}|x_i(k)-x|\leq\max_{1\leq i\leq
n_1}|x_i(0)-x|+L\sum^{k-1}_{s=0}\bar \alpha_s$ and then, for each
$k$,
$$
\bar \alpha_k\max_{1\leq i\leq n_1}|x_i(k)-x|\leq\bar \alpha_k\max_{1\leq i\leq n_1}|x_i(0)-x|+\bar \alpha_k\sum^{k-1}_{s=0}\bar \alpha_sL.
$$
Taking the limit over both sides of the preceding inequality yields the
conclusion. \hfill$\square$

We assume without loss of generality that $m_1=1$ in the sequel of this
subsection for notational simplicity. Denote
$x(k)=(x_1(k),\dots,x_{n_1}(k))'$, $\epsilon(k)=(\epsilon_1(k),\dots,\epsilon_{n_1}(k))'$. Then
system (\ref{0con}) can be written in a compact form:
$$x(k+1)=A_1(k)x(k)+\epsilon(k),k\geq0.
$$
Recall transition matrix
$$\Phi^\ell(k,s)=A_\ell(k)A_\ell(k-1)\cdots A_\ell(s), k\geq s, \ell=1,2.$$
Therefore, for each $k$,
\begin{equation}
\label{con3}x(k+1)=\Phi^1(k,s)x(s)+\sum^{k-1}_{r=s}\Phi^1(k,r+1)\epsilon(r)+\epsilon(k).
\end{equation}

At the end of this section, we present three lemmas for (\ref{6})
(or (\ref{0con}) and the other one for $y$).
The first lemma gives an estimation for
$h_1(k)=\max_{1\leq i,j\leq n_1}|x_i(k)-x_j(k)|$
and $h_2(k)=\max_{1\leq i,j\leq n_2}|y_i(k)-y_j(k)|$
over a bounded interval.

\begin{lem}
\label{con1} Suppose {\bf A2} (ii), {\bf A3} and {\bf A4} hold. Then
for $\ell=1,2$ and any $t\geq1,0\leq q\leq T^\ell-1$,
\begin{align}\label{con4}
h_\ell(t T^\ell+q)&\leq (1-\eta^{T^\ell})h_\ell((t-1)T^\ell+q)\nonumber\\
&\qquad\qquad+2L\sum^{tT^\ell+q-1}_{r=(t-1)T^\ell+q}\lambda^\ell_r,
\end{align}
where $\lambda^1_r=\bar \alpha_r$, $\lambda^2_r=\bar \beta_r$,
$T^\ell=(n_\ell(n_\ell-2)+1)T_\ell$ for a constant $T_\ell$ given in {\bf A2} and
$L$ as the upper bound on the subgradients of objective functions in {\bf A4}.
\end{lem}
\emph{Proof:}
Here we only show the case of $\ell=1$ since the other one can be proven in the same way.
Consider $n_1(n_1-2)+1$ time intervals $[0, T_1-1], [T_1,
2T_1-1],...,[n_1(n_1-2)T_1, (n_1(n_1-2)+1)T_1-1]$. By the definition of
UJSC graph, $\mathcal{G}_1\big([tT_1,(t+1)T_1-1]\big)$ contains a
root node for $0\leq t\leq n_1(n_1-2)$. Clearly, the set of the $n_1(n_1-2)+1$
root nodes contains at least one node, say $i_0$, at least $n_1-1$
times. Assume without loss of generality that $i_0$ is a root node of
$\mathcal{G}_1\big([tT_1,(t+1)T_1-1]\big),t=t_0,...,t_{n_1-2}$.

Take $j_0\neq i_0$ from $\mathcal{V}_1$. It is not hard to show that
there exist a node set $\{j_1,...,j_q\}$ and time set
$\{k_0,...,k_q\}, q\leq n_1-2$ such that $(j_{r+1}, j_r)\in
\mathcal{E}_1(k_{q-r}), 0\leq r\leq q-1$ and $(i_0,j_q)\in
\mathcal{E}_1(k_0),$ where $k_0<\cdots<k_{q-1}<k_q$ and all $k_r$
belong to different intervals $[t_rT_1, (t_r+1)T_1-1], 0\leq r\leq
n_1-2$.

Noticing that the diagonal elements of all adjacency matrices are
positive, and moreover, for matrices $D^1, D^2\in
\mathbb{R}^{n_1\times n_1}$ with nonnegative entries,
$$
(D^1)_{r_0r_1}>0,(D^2)_{r_1r_2}>0 \Longrightarrow (D^1D^2)_{r_0r_2}>0,
$$
so we have $\Phi^1(T^1-1,0)_{j_0i_0}>0$.
Because $j_0$ is taken from $\mathcal{V}_1$ freely,
$\Phi^1(T^1-1,0)_{ji_0}>0$ for $j\in\mathcal{V}_1.$  As a result,
$\Phi^1(T^1-1,0)_{ji_0}\geq\eta^{T^1}$ for $j\in\mathcal{V}_1$ with {\bf
A3} (i) and so $\mu(\Phi^1(T^1-1,0))\leq 1-\eta^{T^1}$ by the
definition of ergodicity coefficient given in Lemma \ref{cons}.
According to (\ref{con3}), the inequality $\hbar(\mu+\nu)\leq \hbar(\mu)+2\max_i\nu_i$,
Lemma \ref{cons} and {\bf A4},
\begin{align}
h_1(T^1)&\leq h_1(\Phi^1(T^1-1,0)x(0))+2L\sum^{T^1-1}_{r=0}\bar \alpha_r\nonumber\\
&\leq \mu(\Phi^1(T^1-1,0))h_1(0)+2L\sum^{T^1-1}_{r=0}\bar \alpha_r\nonumber\\
&\leq (1-\eta^{T^1}) h_1(0)+2L\sum^{T^1-1}_{r=0}\bar \alpha_r,\nonumber
\end{align}
which shows (\ref{con4}) for $\ell=1,t=1,q=0$.
Analogously, we can show
(\ref{con4}) for $\ell=1,2$ and $t\geq1,0\leq q\leq T^\ell-1$. \hfill$\square$

\begin{lem}
\label{co} Suppose {\bf A2} (ii), {\bf A3} and {\bf A4} hold.

(i) If $\sum^{\infty}_{k=0}\bar \alpha^2_k<\infty$
and $\sum^{\infty}_{k=0}\bar \beta^2_k<\infty$, then
$\sum^\infty_{k=0}\bar \alpha_kh_1(k)<\infty,$ $\sum^\infty_{k=0}\bar \beta_kh_2(k)<\infty;$

(ii) If for each $i$, $\lim_{k\to\infty}\alpha_{i,k}=0$
and $\lim_{k\to\infty}\beta_{i,k}=0$, then the subnetworks $\Xi_1$ and $\Xi_2$ achieve a consensus, respectively, i.e., $\lim_{k\to\infty}h_1(k)=0,\lim_{k\to\infty}h_2(k)=0.$
\end{lem}

Note that (i) is an extension of Lemma 8 (b) in
\cite{nedic2} dealing with weight-balanced graph sequence to
general graph sequence (possibly weight-unbalanced),
while (ii) is about the consensus within the subnetworks, and will be frequently used in the sequel.
This lemma can be shown by Lemma \ref{con1} and similar arguments to
the proof of Lemma 8 in \cite{nedic2}, and hence, the proof is omitted here.

The following provides the error estimation between
agents' states and their average.
\begin{lem}
\label{con0} Suppose {\bf A2}--{\bf A4} hold, and
$\{\bar \alpha(k)\}$, $\{\bar \beta(k)\}$ are non-increasing with $\sum^{\infty}_{k=0}\bar \alpha^2_k<\infty$, $\sum^{\infty}_{k=0}\bar \beta^2_k<\infty$. Then for each $i\in \mathcal{V}_1$
and $j\in \mathcal{V}_2$,
$\sum^\infty_{k=0}\bar \beta_k|\breve{x}_i(k)-\bar y(k)|<\infty$,
$\sum^\infty_{k=0}\bar \alpha_k|\breve{y}_j(k)-\bar x(k)|<\infty,$
where
$\bar x(k)=\frac{1}{n_1}\sum^{n_1}_{i=1}x_i(k)\in X$,
$\bar y(k)=\frac{1}{n_2}\sum^{n_2}_{i=1}y_i(k)\in Y$.
\end{lem}

\vskip 3mm
\emph{Proof:}
We only need to show the first conclusion since the second one
can be obtained in the same way.
At first, from {\bf A3} (iii)
and $|y_j(\breve{k}_i)-\bar y(\breve{k}_i)|\leq h_2(\breve{k}_i)$ we have
\begin{align}\label{con11}
&\sum^\infty_{k=0}\bar \beta_k|\breve{x}_i(k)-\bar y(\breve{k}_i)|\leq\sum^\infty_{k=0}\bar \beta_kh_2(\breve{k}_i).
\end{align}
Let $\{s_{ir},r\geq0\}$ be the set of all moments when
$\mathcal{N}^{2}_i(s_{ir})\neq\emptyset$.  Recalling the definition of $\breve{k}_i$
in (\ref{8}), $\breve{k}_i=s_{ir}$ when $s_{ir}\leq k<s_{i(r+1)}$.
Since $\{\bar \beta_k\}$ is non-increasing and $\sum^\infty_{k=0}
\bar \beta_kh_2(k)<\infty$ (by Lemma \ref{co}), we have
\begin{align}
&\sum^\infty_{k=0}\bar \beta_kh_2(\breve{k}_i)\leq \sum^\infty_{k=0}\bar \beta_{\breve{k}_i}h_2(\breve{k}_i)\nonumber\\
&=\sum^\infty_{r=0}\bar \beta_{s_{ir}}|s_{i(r+1)}-s_{ir}|h_2(s_{ir})\nonumber\\
&\leq T_{\bowtie}\sum^\infty_{r=0} \bar \beta_{s_{ir}}h_2(s_{ir})\leq T_{\bowtie}\sum^\infty_{k=0} \bar \beta_{k}h_2(k)<\infty,\nonumber
\end{align}
where $T_{\bowtie}$ is the constant in {\bf A2} (i). Thus, the preceding inequality and (\ref{con11})
imply $\sum^\infty_{k=0}\bar \beta_k|\breve{x}_i(k)-\bar
y(\breve{k}_i)|<\infty$.

Since $y_i(k)\in Y$ for all $i$ and $Y$ is convex, $\bar y(k)\in Y$.
Then, from the non-expansiveness property of the convex projection operator,
\begin{align}\label{con12}
&|\bar y(k+1)-\bar y(k)|\nonumber\\
&=\bigg|\frac{\sum^{n_2}_{i=1}
\big(P_Y(\hat y_i(k)+\beta_{i,k}q_{2i}(k))-P_Y(\bar y(k))\big)}{n_2}\bigg|\nonumber\\
&\leq \frac{1}{n_2}\sum^{n_2}_{i=1}\big|\hat y_i(k)+\beta_{i,k}q_{2i}(k)-\bar y(k)\big|\nonumber\\
&\leq h_2(k)+\bar\beta_kL.
\end{align}
Based on (\ref{con12}), the non-increasingness of $\{\bar \beta_k\}$ and $\breve{k}_i\geq k-T_{\bowtie}+1$, we
also have
\begin{align}
&\sum^\infty_{k=0}\bar \beta_k|\bar y(\breve{k}_i)-\bar y(k)|\leq\sum^\infty_{k=0}\bar \beta_k\sum^{k-1}_{r=\breve{k}_i}\big|\bar y(r)-\bar y(r+1)\big|\nonumber\\
&\leq\sum^\infty_{k=0}\bar \beta_k\sum^{k-1}_{r=\breve{k}_i}(h_2(r)+\bar \beta_rL)\nonumber\\
&\leq\sum^\infty_{k=0}\bar \beta_k\sum^{k-1}_{r=k-T_{\bowtie}+1}(h_2(r)+\bar \beta_rL)\nonumber\\
&\leq\sum^\infty_{k=0}\bar \beta_k\sum^{k-1}_{r=k-T_{\bowtie}+1}h_2(r)+\frac{(T_{\bowtie}-1)L}{2}\sum^\infty_{k=0}\bar \beta^2_k\nonumber\\
&\quad+\frac{L}{2}\sum^\infty_{k=0}\sum^{k-1}_{r=k-T_{\bowtie}+1}\bar \beta^2_r\nonumber\\
&\leq(T_{\bowtie}-1)\sum^\infty_{k=0}\bar \beta_kh_2(k)+\frac{(T_{\bowtie}-1)L}{2}\sum^\infty_{k=0}\bar \beta^2_k\nonumber\\
&\quad+\frac{(T_{\bowtie}-1)L}{2}\sum^\infty_{k=0}\bar \beta^2_k<\infty,\nonumber
\end{align}
where $h_2(r)=\bar\beta_r=0$, $r<0$.
Since $|\breve{x}_i(k)-\bar y(k)|\leq |\breve{x}_i(k)-\bar
y(\breve{k}_i)|+|\bar y(\breve{k}_i)-\bar y(k)|$, the first conclusion
follows. \hfill$\square$

\begin{rem}\label{rem}
From the proof we find that Lemma \ref{con0} still holds when
the non-increasing condition of $\{\bar \alpha_k\}$ and $\{\bar
\beta_k\}$ is replaced by that there are an integer $T^*>0$
and $c^*>0$ such that $\bar \alpha_{k+T^*}\leq c^*\bar
\alpha_k$ and $\bar \beta_{k+T^*}\leq c^*\bar \beta_k$ for all $k$.
\end{rem}

\subsection{Proof of Theorem \ref{thm1}}

We complete the proof by the following two steps.

\emph{Step 1:} We first show that the states of (\ref{6}) are bounded.
Take $(x, y)\in X\times Y$. By (\ref{6}) and (\ref{pro}),
\begin{align}\label{th1}
&|x_i(k+1)-x|^2\leq|\hat x_i(k)-\gamma_kq_{1i}(k)-x|^2=|\hat x_i(k)-x|^2\nonumber\\
&\qquad+2\gamma_k\big\langle \hat x_i(k)-x, -q_{1i}(k) \big\rangle+\gamma^2_k|q_{1i}(k)|^2.
\end{align}

It is easy to see that $|\cdot|^2$ is a convex function from the convexity of $|\cdot|$ and the convexity of scalar function $h(c)=c^2$. From this and {\bf A3} (ii), $|\hat x_i(k)-x|^2\leq
\sum_{j\in\mathcal{N}^1_i(k)}a_{ij}(k)|x_j(k)-x|^2$. Moreover,
since $q_{1i}(k)$ is a subgradient of $f_i(\cdot,\breve{x}_i(k))$ at $\hat x_i(k)$,
$\langle x-\hat x_i(k), q_{1i}(k) \rangle\leq
f_{i}(x,\breve{x}_i(k))- f_{i}(\hat x_i(k),\breve{x}_i(k)).$
 Thus, based on
(\ref{th1}) and {\bf A4},
\begin{align}\label{th2}
&|x_i(k+1)-x|^2\leq\sum_{j\in\mathcal{N}^1_i(k)}a_{ij}(k)|x_j(k)-x|^2+L^2\gamma^2_k\nonumber\\
&\qquad\qquad+2\gamma_k\big(f_{i}(x,
 \breve{x}_i(k))- f_{i}(\hat x_i(k),
 \breve{x}_i(k))\big).
\end{align}
Again employing {\bf A4},
$|f_{i}(x,y_1)-f_{i}(x,y_2)|\leq L|y_1-y_2|,$
$|f_{i}(x_1,y)-f_{i}(x_2,y)|\leq L|x_1-x_2|, \forall
x,x_1,x_2\in X,y,y_1,y_2\in Y$.
This imply
\begin{align}
&|f_{i}\big(x,\breve{x}_i(k)\big)-f_{i}\big(x,\bar y(k)\big)|\leq L|\breve{x}_i(k)-\bar y(k)|,\label{th3}\\
&\big|f_{i}\big(\hat x_i(k),\breve{x}_i(k)-f_{i}\big(\bar x(k),\bar
y(k)\big)\big|\nonumber\\
&\leq L\big(|\hat x_i(k)-\bar x(k)|+|\breve{x}_i(k)-\bar y(k)|\big)\nonumber\\
\label{th4}&\leq L\big(h_1(k)+|\breve{x}_i(k)-\bar y(k)|\big).
\end{align}
Hence, by (\ref{th2}), (\ref{th3}) and (\ref{th4}),
\begin{align}\label{th5}
&|x_i(k+1)-x|^2\leq\sum_{j\in\mathcal{N}^1_i(k)}a_{ij}(k)|x_j(k)-x|^2\nonumber\\
&+2\gamma_k(f_{i}(x,\bar y(k)\big)-f_{i}\big(\bar x(k),
 \bar y(k)))+L^2\gamma^2_k+2L\gamma_ke_{i1}(k),
\end{align}
where $e_{i1}(k)=h_1(k)+2|\breve{x}_i(k)-\bar y(k)|$.

It follows from the weight balance of $\mathcal{G}_1(k)$ and {\bf A3} (ii) that
$\sum_{i\in\mathcal{V}_1}a_{ij}(k)=1$ for all $j\in\mathcal{V}_1$.
Then, from
(\ref{th5}), we have
\begin{align}\label{th6}
&\sum^{n_1}_{i=1}|x_i(k+1)-x|^2\leq\sum^{n_1}_{i=1}|x_i(k)-x|^2+2\gamma_k\big(U(x,\bar y(k))\nonumber\\
&\quad- U(\bar x(k),
 \bar y(k))\big)+n_1L^2\gamma^2_k+2L\gamma_k\sum^{n_1}_{i=1}e_{i1}(k).
\end{align}
Analogously,
\begin{align}\label{th7}
&\sum^{n_2}_{i=1}|y_i(k+1)-y|^2\leq\sum^{n_2}_{i=1}|y_i(k)-y|^2+
2\gamma_k(U(\bar x(k),\bar y(k))\nonumber\\
&\quad-U(\bar x(k),y))+n_2L^2\gamma^2_k+2L\gamma_k\sum^{n_2}_{i=1}e_{i2}(k),
\end{align}
where $e_{i2}(k)=h_2(k)+2|\breve{y}_i(k)-\bar x(k)|$.
Let $(x,y)=(x^*,y^*)\in X^*\times Y^*$, which is nonempty by {\bf A1}.
Denote $\xi(k, x^*,y^*)=\sum^{n_1}_{i=1}|x_i(k)-x^*|^2+\sum^{n_2}_{i=1}|y_i(k)-y^*|^2$.
Then adding (\ref{th6}) and (\ref{th7}) together leads to
\begin{align}\label{th8}
&\xi(k+1, x^*, y^*)\leq\xi(k, x^*, y^*)-2\gamma_k\Upsilon(k)\nonumber\\
&\quad+(n_1+n_2)L^2\gamma^2_k+2L\gamma_k\sum^{2}_{\ell=1}\sum^{n_\ell}_{i=1}e_{i\ell}(k),
\end{align}
where
\begin{align}\label{sp}
\Upsilon(k)&=U(\bar x(k),y^*)-U(x^*,\bar y(k))\nonumber\\
&=U(x^*, y^*)-U(x^*,\bar y(k))+U(\bar x(k),y^*)-U(x^*, y^*)\nonumber\\
&\geq0
\end{align}
following from $U(x^*,y^*)-U(x^*,\bar y(k))\geq0$, $U(\bar
x(k),y^*)-U(x^*,y^*)\geq0$ for $k\geq 0$
since $(x^*,y^*)$ is a saddle point of $U$ on $X\times Y$.
Moreover, by $\sum^{\infty}_{k=0}\gamma^2_k<\infty$ and Lemmas \ref{co}, \ref{con0},
\begin{align}\label{2v}
\sum^{\infty}_{k=0}\gamma_k\sum^{2}_{\ell=1}\sum^{n_\ell}_{i=1}e_{i\ell}(k)<\infty.
\end{align}
Therefore, by virtue of $\sum^{\infty}_{k=0}\gamma^2_k<\infty$ again, (\ref{2v}), (\ref{th8}) and Lemma
\ref{4.1}, $\lim_{k\to\infty}\xi(k,x^*,y^*)$ is a finite number,
denoted as $\xi(x^*,y^*)$. Thus, the conclusion follows.
\hfill$\square$

\emph{Step 2:} We next
show that the limit points of all agents satisfy certain
objective function equations, and then prove the Nash equilibrium convergence
 under either of the two conditions: (i) and (ii).

As shows in \emph{Step 1}, $(x_i(k),y_i(k)),k\geq0$ are bounded.   Moreover, it also follows from (\ref{th8})
that
\begin{align}
2\sum^{k}_{r=0}\gamma_r\Upsilon(r)&\leq \xi(0,x^*,y^*)+(n_1+n_2)L^2\sum^{k}_{r=0}\gamma^2_r\nonumber\\
&+2L\sum^{k}_{r=0}\gamma_r
\sum^{2}_{\ell=1}\sum^{n_\ell}_{i=1}e_{i\ell}(r)\nonumber
\end{align}
and then by $\sum^{\infty}_{k=0}\gamma^2_k<\infty$ and (\ref{2v}) we have
\begin{align}\label{th10}
0\leq&\sum^{\infty}_{k=0}\gamma_k\Upsilon(k)<\infty.
\end{align}

The stepsize condition $\sum^{\infty}_{k=0}\gamma_k=\infty$ and
(\ref{th10}) imply $\liminf_{k\rightarrow\infty}\Upsilon(k)=0.$
As a result, there is a subsequence $\{k_r\}$ such that
$\lim_{r\to\infty}U(x^*,\bar y(k_r))=U(x^*,y^*)$
and $\lim_{r\to\infty}U(\bar x(k_r),y^*)=U(x^*,y^*)$. Let
$(\tilde{x},\tilde{y})$ be any limit pair of $\{(\bar x(k_r),\bar y(k_r))\}$ (noting that the finite limit pairs exist
by the boundedness of system states).  Because $U(x^*,\cdot),U(\cdot,y^*)$ are
continuous and the Nash equilibrium point $(x^*,y^*)$ is taken from $X^*\times
Y^*$ freely, the limit pair $(\tilde{x},\tilde{y})$ must satisfy
that for any $(x^*,y^*)\in X^*\times Y^*$,
\begin{align}\label{th11}
U(x^*,\tilde{y})=U(\tilde{x},y^*)=U(x^*,y^*).
\end{align}

We complete the proof by discussing the proposed two sufficient
conditions: (i) and (ii).

(i). For the strictly convex-concave function $U$, we claim that
$X^*\times Y^*$ is a single-point set. If it contains two different
points $(x^*_1,y^*_1)$ and $(x^*_2,y^*_2)$ (without loss of
generality, assume $x^*_1\neq x^*_2$), it also contains
point $(x^*_2,y^*_1)$ by Lemma \ref{lem01}. Thus, $U(x^*_1,y^*_1)\leq U(x,y^*_1)$
and $U(x^*_2,y^*_1)\leq U(x,y^*_1)$ for any $x\in X$, which yields a contradiction since
$U(\cdot,y^*_1)$ is strictly convex and then the minimizer of
$U(\cdot,y^*_1)$ is unique. Thus, $X^*\times Y^*$ contains only one
single-point (denoted as $(x^*,y^*)$).

Then $\tilde{x}=x^*,\tilde{y}=y^*$ by (\ref{th11}).
Consequently, each limit pair of $\{(\bar x(k_r),\bar y(k_r))\}$ is $(x^*,y^*)$, i.e.,
$\lim_{r\rightarrow\infty}\bar x(k_r)=x^*$ and
$\lim_{r\rightarrow\infty}\bar y(k_r)=y^*$. By Lemma
\ref{co}, $\lim_{r\rightarrow\infty}x_i(k_r)=x^*$, $i\in
\mathcal{V}_1$ and $\lim_{r\rightarrow\infty}y_i(k_r)=y^*$, $i\in
\mathcal{V}_2$. Moreover,
$\lim_{k\rightarrow\infty}\xi(k,x^*,y^*)=\xi(x^*,y^*)$ as given in
\emph{Step 1}, so
$\xi(x^*,y^*)=\lim_{r\rightarrow\infty}\xi(k_r,x^*,y^*)=0$, which in
return implies $\lim_{k\rightarrow\infty} x_i(k)$ $=x^*$, $i\in
\mathcal{V}_1$ and $\lim_{k\rightarrow\infty} y_i(k)=y^*$, $i\in
\mathcal{V}_2$.

(ii). In \emph{Step 1}, we proved
$\lim_{k\rightarrow\infty}\xi(k,x^*,y^*)=\xi(x^*,y^*)$ for any
$(x^*, y^*)\in X^*\times Y^*$. We check the existence of the two
limits $\lim_{k\rightarrow\infty}\bar x(k)$ and
$\lim_{k\rightarrow\infty}\bar y(k)$. Let $(x^+,y^+)$ be an interior
point of $X^*\times Y^*$ for which
$\mathbf{B}(x^+,\varepsilon)\subseteq X^*$ and
$\mathbf{B}(y^+,\varepsilon)\subseteq Y^*$ for some $\varepsilon>0$.
Clearly, any two limit pairs $(\grave{x}_1, \grave{y}_1)$,
$(\grave{x}_2, \grave{y}_2)$ of $\{(\bar x(k),\bar y(k))\}$
must satisfy
$n_1|\grave{x}_1-x|^2+n_2|\grave{y}_1-y|^2=n_1|\grave{x}_2-x|^2+n_2|\grave{y}_2-y|^2,
\;\;\forall x\in \mathbf{B}(x^+,\varepsilon),\;y\in
\mathbf{B}(y^+,\varepsilon).$
Take $y=y^+$. Then for any $x\in \mathbf{B}(x^+,\varepsilon)$,
\begin{align}\label{th13}
n_1|\grave{x}_1-x|^2=n_1|\grave{x}_2-x|^2+n_2\big(|\grave{y}_2-y^+|^2-|\grave{y}_1-y^+|^2\big).
\end{align}
Taking the gradient with respect to $x$ on both sides of
(\ref{th13}) yields $2n_1(x-\grave{x}_1)=2n_1(x-\grave{x}_2)$,
namely, $\grave{x}_1=\grave{x}_2$. Similarly, we can show
$\grave{y}_1=\grave{y}_2$. Thus, the limits,
$\lim_{k\rightarrow\infty}\bar x(k)=\grave{x}\in X$ and
$\lim_{k\rightarrow\infty}\bar y(k)=\grave{y}\in Y$, exist.
Based on Lemma \ref{co} (ii), $\lim_{k\rightarrow\infty}
x_i(k)=\grave{x}, i\in \mathcal{V}_1$ and $\lim_{k\rightarrow\infty}
y_i(k)=\grave{y}, i\in \mathcal{V}_2$.

We claim that $(\grave{x},
\grave{y})\in X^*\times Y^*$. First it follows from (\ref{th6}) that, for
any $x\in X$,
$\sum^\infty_{k=0}\gamma_k\sum^{n_1}_{i=1}\big(U(\bar x(k),\bar y(k))-U(x,\bar y(k))\big)<\infty.$
Moreover, recalling $\sum^\infty_{k=0}\gamma_k=\infty$, we obtain
\begin{align}\label{add1}
\liminf_{k\rightarrow\infty}\big(U(\bar x(k),\bar y(k))-U(x,\bar y(k))\big)\leq0.
\end{align}
Then $U(\grave{x},\grave{y})-U(x,\grave{y})\leq0$ for all $x\in X$ due to
$\lim_{k\rightarrow\infty}\bar x(k)=\grave{x}$,
$\lim_{k\rightarrow\infty}\bar y(k)=\grave{y}$, the continuity of $U$, and (\ref{add1}).
Similarly, we can show
$U(\grave{x},y)-U(\grave{x},\grave{y})\leq0$ for all $y\in Y$. Thus, $(\grave{x},\grave{y})$ is
a saddle point of $U$ on $X\times Y$, which implies
$(\grave{x},\grave{y})\in X^*\times Y^*$.

Thus, the proof is
completed. \hfill$\square$

\subsection{Proof of Theorem \ref{thmdao1}}

 (Necessity) Let $(x^*,y^*)$ be the unique saddle point
of strictly convex-concave function $U$ on $X\times Y$. Take
$\mu=(\mu_1,\dots,\mu_{n_1})'\in\mathcal{S}^+_{n_1}$. By Lemma
\ref{4.3} again, there is a stochastic matrix $A_1$ such that
$\mu'A_1=\mu'$ and $\mathcal{G}_{A_1}$ is strongly connected.
Let $\mathcal{G}_1=\{\mathcal{G}_1(k)\}$ be the graph sequence of algorithm (\ref{6}) with
$\mathcal{G}_1(k)=\mathcal{G}_{A_1}$ for $k\geq 0$
and $A_1$ being the adjacency matrix of $\mathcal{G}_1(k)$.
Clearly, $\mathcal{G}_1$ is UJSC.
 On one hand, by Proposition  \ref{th6.3}, all agents converge to the
unique saddle point of
$\sum^{n_1}_{i=1}\mu_if_i$ on $X\times Y$. On the other hand, the
necessity condition states that
$\lim_{k\rightarrow\infty}x_i(k)=x^*$ and
$\lim_{k\rightarrow\infty}y_i(k)=y^*$ for $i=1,...,n_1$. Therefore,
$(x^*,y^*)$ is a saddle point of
$\sum^{n_1}_{i=1}\mu_if_i$ on $X\times Y$.

Because $\mu$ is taken from $\mathcal{S}^+_{n_1}$ freely, we have
that, for any $\mu\in\mathcal{S}^+_{n_1}$, $x\in X$, $y\in Y$,
\begin{align}\label{do90}
\sum^{n_1}_{i=1}\mu_if_i(x^*,y)\leq\sum^{n_1}_{i=1}\mu_if_i(x^*,y^*)
\leq\sum^{n_1}_{i=1}\mu_if_i(x,y^*).
\end{align}
We next show by contradiction that, given any $i=1,...,n_1$,
$f_i(x^*,y^*)\leq f_i(x,y^*)$ for all $x\in X$. Hence suppose there are $i_0$
and $\hat x\in X$ such that $f_{i_0}(x^*,y^*)>f_{i_0}(\hat x,y^*)$. Let
$\mu_i,i\neq i_0$ be sufficiently small such that $\big|\sum_{i\neq
i_0}\mu_if_i(x^*,y^*)\big|<\frac{\mu_{i_0}}{2}(f_{i_0}(x^*,y^*)-f_{i_0}(\hat x,y^*))$
and $\big|\sum_{i\neq i_0}\mu_if_i(\hat x,y^*)\big|<\frac{\mu_{i_0}}{2}(f_{i_0}(x^*,y^*)-f_{i_0}(\hat x,y^*))$.
Consequently,
$\sum^{n_1}_{i=1}\mu_if_i(x^*,y^*)>\frac{\mu_{i_0}}{2}\big(f_{i_0}(x^*,y^*)+f_{i_0}(\hat x,y^*)\big)>\sum^{n_1}_{i=1}\mu_if_i(\hat x,y^*),$
which contradicts the second inequality of
(\ref{do90}). Thus, $f_i(x^*,y^*)\leq f_i(x,y^*)$ for all $x\in X$. Analogously, we can show from the first inequality of (\ref{do90}) that for each $i$, $f_i(x^*,y)\leq f_i(x^*,y^*)$ for all $y\in Y$. Thus, we obtain that
$f_i(x^*,y)\leq f_i(x^*,y^*)
\leq f_i(x,y^*)$, $\forall x\in X, y\in Y,$
or equivalently, $(x^*,y^*)$ is the saddle point of
$f_i$ on $X\times Y$.

(Sufficiency) Let $(x^*,y^*)$ be the unique saddle point of
$f_i,i=1,...,n_1$ on $X\times Y$. Similar to (\ref{th5}), we have
\begin{align}\label{dao12}
&|y_i(k+1)-y^*|^2\leq\sum_{j\in\mathcal{N}^1_i(k)}a_{ij}(k)|y_j(k)-y^*|^2\nonumber\\
&+2\gamma_k\big(f_i(\bar x(k),
 \bar y(k))-f_i(\bar x(k), y^*)\big)+L^2\gamma^2_k+2L\gamma_ku_{2}(k),
\end{align}
where $u_{2}(k)=2h_1(k)+h_2(k)$.
Merging (\ref{th5}) and (\ref{dao12}) gives
\begin{align}\label{dao1}
&\zeta(k+1)\leq \zeta(k)+
2\gamma_k\max_{1\leq i\leq n_1}(f_i(x^*,\bar y(k))- f_i(\bar x(k),
 y^*))\nonumber\\
&\qquad\qquad+2L^2\gamma^2_k+2L\gamma_k(u_{1}(k)+u_{2}(k))\nonumber\\
&=\zeta(k)+2\gamma_k\max_{1\leq i\leq n_1}(f_i(x^*,\bar y(k))-f_i(x^*,y^*)\nonumber\\
&\qquad\qquad+f_i(x^*,y^*)-f_i(\bar x(k),y^*))
+2L^2\gamma^2_k\nonumber\\
&\qquad\qquad+6L\gamma_k(h_{1}(k)+h_{2}(k)),
\end{align}
where $\zeta(k)=\max_{1\leq i\leq
n_1}(|x_i(k)-x^*|^2+|y_i(k)-y^*|^2)$,
$u_{1}(k)=h_1(k)+2h_2(k)$. Since
$f_i(x^*,\bar y(k))-f_i(x^*,y^*)\leq 0$ and
$f_i(x^*,y^*)- f_i(\bar x(k),y^*)\leq0$ for all
$i,k$, the second term in (\ref{dao1}) is non-positive. By Lemma
\ref{4.1},
\begin{equation}
\label{kpsi} \lim_{k\to\infty}\zeta(k)=\zeta^*\geq 0
\end{equation}
for a finite number $\zeta^*$, which implies that $(x_i(k),y_i(k)),k\geq 0$ are bounded.

Denote $\wp(k)=\min_{1\leq i\leq n_1}(f_i(x^*,y^*)
-f_i(x^*,\bar y(k))+ f_i(\bar x(k),y^*)-f_i(x^*,y^*))$. From (\ref{dao1}), we also have
\begin{align}
0&\leq2\sum^k_{l=0}\gamma_l\wp(l)\leq\zeta(0)-\zeta(k+1)+2L^2\sum^k_{l=0}\gamma^2_l\nonumber\\
 &\qquad+6L\sum^k_{l=0}\gamma_l(h_{1}(l)+h_{2}(l)),\; k\geq 0,\nonumber
\end{align}
and hence $0\leq\sum^\infty_{k=0}\gamma_k\wp(k)<\infty.$
The stepsize condition $\sum^\infty_{k=0}\gamma_k=\infty$ implies
that there is a subsequence $\{k_r\}$ such that
$$
\lim_{r\to\infty}\wp(k_r)=0.
$$
We assume without loss of generality that $\lim_{r\to\infty}\bar
x(k_r)=\acute{x}, \lim_{r\to\infty}\bar y(k_r)=\acute{y}$ for some
$\acute{x},\acute{y}$ (otherwise we can find a subsequence of
$\{k_r\}$ recalling the boundedness of system
states). Due to the finite number of agents and the
continuity of $f_i$s, there exists $i_0$ such
that $f_{i_0}(x^*,y^*)=f_{i_0}(x^*,\acute{y})$ and
$f_{i_0}(\acute{x},y^*)=f_{i_0}(x^*,y^*)$. It
follows from the strict convexity-concavity of $f_{i_0}$ that
$\acute{x}=x^*,\acute{y}=y^*$.

Since the consensus is achieved within two subnetworks,
$\lim_{r\to\infty}x_i(k_r)=x^*$ and $\lim_{r\to\infty}y_i(k_r)=y^*$, which leads to
$\zeta^*=0$ based on (\ref{kpsi}). Thus, the conclusion follows.
\hfill$\square$

\subsection{Proof of Theorem \ref{th4.4}}

We design the stepsizes $\alpha_{i,k}$ and $\beta_{i,k}$
as that given before Remark \ref{rem4.4}.
First by Lemma \ref{lem1} (i) and (ii), the limit
$\lim_{r\rightarrow\infty}\Phi^\ell(r,k)=\textbf{1}(\phi^\ell(k))'$
exists for each $k$.
Let $(x^*,y^*)$ be the unique Nash equilibrium. From (\ref{th5}) we have
\begin{align}\label{ex1}
&|x_i(k+1)-x^*|^2\leq\sum_{j\in\mathcal{N}^1_i(k)}a_{ij}(k)|x_j(k)-x^*|^2\nonumber\\
&\qquad+2\alpha_{i,k}(f_{i}(x^*,\bar y(k))- f_{i}(\bar x(k),
 \bar y(k)))\nonumber\\
&\qquad+L^2\alpha^2_{i,k}+2L\alpha_{i,k}e_{i1}(k).
\end{align}
Analogously,
\begin{align}\label{ex0}
&|y_i(k+1)-y^*|^2\leq\sum_{j\in\mathcal{N}^2_i(k)}a_{ij}(k)|y_j(k)-y^*|^2\nonumber\\
&\qquad+2\beta_{i,k}\big(g_{i}(\bar x(k),\bar y(k))-g_{i}(\bar x(k),y^*)\big)\nonumber\\
&\qquad+L^2\beta^2_{i,k}+2L\beta_{i,k}e_{i2}(k).
\end{align}
Denote
\begin{align}
&\Lambda^1_k=\mbox{diag}\Big\{\frac{1}{\alpha^1_k},\dots,\frac{1}{\alpha^{n_1}_k}\Big\},
\Lambda^2_k=\mbox{diag}\Big\{\frac{1}{\beta^1_k},\dots,\frac{1}{\beta^{n_2}_k}\Big\};\nonumber\\
&\psi^\ell(k)=(\psi^\ell_1(k),\dots,\psi^\ell_{n_\ell}(k))', \ell=1,2,\nonumber\\
&\qquad\psi^1_i(k)=|x_i(k)-x^*|^2, \psi^2_i(k)=|y_i(k)-y^*|^2;\nonumber\\
&\vartheta^\ell(k)=(\vartheta^\ell_1(k),\dots,\vartheta^\ell_{n_\ell}(k))',\nonumber\\
&\qquad\vartheta^1_i(k)=f_{i}(\bar x(k),\bar y(k))-f_{i}(x^*,\bar y(k)),\nonumber\\
&\qquad\vartheta^2_i(k)=g_{i}(\bar x(k),y^*)-g_{i}(\bar x(k),\bar y(k));\nonumber\\
&e_\ell(k)=(e_{1\ell}(k),\dots, e_{n_\ell\ell}(k))'.\nonumber
\end{align}
 Then it follows from (\ref{ex1}) and (\ref{ex0}) that
\begin{align}
\psi^\ell(k+1)
&\leq A_\ell(k)\psi^\ell(k)-2\gamma_k\Lambda^\ell_k\vartheta^\ell(k)\nonumber\\
&\qquad+\delta^2_*L^2\gamma^2_k\textbf{1}+2\delta_*L\gamma_ke_\ell(k),\nonumber
\end{align}
where $\delta_*=\sup_{i,k}\{1/\alpha^i_k,1/\beta^i_k\}$.
By Lemma \ref{lem1} (iii),
$\alpha^i_k\geq\eta^{(n_1-1)T_1},\beta^i_k\geq\eta^{(n_2-1)T_2},\forall i,k$
 and then  $\delta_*$ is a finite number.
Therefore,
\begin{align}\label{ex3}
&\psi^\ell(k+1)\leq\Phi^\ell(k,r)\psi^\ell(r)-2\sum^{k-1}_{s=r}\gamma_s
\Phi^\ell(k,s+1)\Lambda^\ell_s\vartheta^\ell(s)\nonumber\\
&\qquad+\delta^2_*L^2\sum^{k}_{s=r}\gamma^2_s\textbf{1}
+2\delta_*L\sum^{k-1}_{s=r}\gamma_s\Phi^\ell(k,s+1)e_\ell(s)\nonumber\\
&\qquad-2\gamma_k\Lambda^\ell_k\vartheta^\ell(k)+2\delta_*L\gamma_ke_\ell(k).
\end{align}
 Then (\ref{ex3}) can be written as
\begin{align}\label{ex4}
&\psi^\ell(k+1)\nonumber\\
&\leq\Phi^\ell(k,r)\psi^\ell(r)
-2\sum^{k-1}_{s=r}\gamma_s\textbf{1}(\phi^\ell(s+1))'\Lambda^\ell_s\vartheta^\ell(s)\nonumber\\
&\;+\delta^2_*L^2\sum^{k}_{s=r}\gamma^2_s\textbf{1}
+2\delta_*L\sum^{k-1}_{s=r}\gamma_s\textbf{1}(\phi^\ell(s+1))'e_\ell(s)\nonumber\\
&\;+2\sum^{k-1}_{s=r}\gamma_s\big(\textbf{1}(\phi^\ell(s+1))'-\Phi^\ell(k,s+1)\big)\Lambda^\ell_s\vartheta^\ell(s)\nonumber\\
&\;-2\gamma_k\Lambda^\ell_k\vartheta^\ell(k)+2\delta_*L\gamma_ke_\ell(k)\nonumber\\
&\;+2\delta_*L\sum^{k-1}_{s=r}\gamma_s\big(\Phi^\ell(k,s+1)-\textbf{1}(\phi^\ell(s+1))'\big)e_\ell(s).
\end{align}

The subsequent proof is given as follows. First, we show that the designed stepsizes (\ref{desi2})
can eliminate the imbalance caused by the weight-unbalanced graphs (see the second term in (\ref{ex4})), and then we prove that all the terms from the third one to the last one in (\ref{ex4}) is summable
based on the geometric rate convergence of transition matrices.
Finally, we show the desired convergence based on inequality (\ref{ex4}),
as (\ref{th8}) for the weight-balance case in Theorem \ref{thm1}.

Clearly,
$\textbf{1}(\phi^\ell(s+1))'\Lambda^\ell_s=\textbf{1}\textbf{1}',\ell=1,2$.
From Lemma \ref{lem1} (iv) we also have that
$\big|\Phi^\ell(k,s)_{ij}-\phi^\ell_j(s)\big|\leq C \rho^{k-s}$ for $\ell=1,2$, every
$i=1,...,n_\ell$, $s\geq 0$, $k\geq s$, and $j=1,...,n_\ell$, where $C=\max\{C_1,C_2\}$,
$0<\rho=\max\{\rho_1,\rho_2\}<1$. Moreover, by {\bf A4}, $|\vartheta^1_i(s)|=|f_{i}(\bar
x(s),\bar y(s))-f_{i}(x^*,\bar y(s))|\leq L|\bar x(s)-x^*|$ for $i\in
\mathcal{V}_1$, and
$|\vartheta^2_i(s)|=|f_{i}(\bar x(s),y^*)-f_{i}(\bar
x(s),\bar y(s))|\leq L|\bar y(s)-y^*|$ for $i\in \mathcal{V}_2$.
Based on these observations,
multiplying $\frac{1}{n_\ell}\textbf{1}'$ on the both sides of (\ref{ex4}) and taking the sum over $\ell=1,2$ yield
\begin{align}
&\sum^2_{\ell=1}\frac{1}{n_\ell}\textbf{1}'\psi^\ell(k+1)
\leq\sum^2_{\ell=1}\frac{1}{n_\ell}\textbf{1}'\Phi^\ell(k,r)\psi^\ell(r)\nonumber\\
&\qquad-2\sum^{k-1}_{s=r}\gamma_s\sum^2_{\ell=1}\sum^{n_\ell}_{i=1}\vartheta^\ell_i(s)
+2\delta^2_*L^2\sum^{k}_{s=r}\gamma^2_s\nonumber\\
&\qquad+2\delta_*L\sum^{k-1}_{s=r}\gamma_s\sum^2_{\ell=1}\sum^{n_\ell}_{i=1}e_{i\ell}(s)\nonumber\\
&\qquad+2CL\delta_*(n_1+n_2)\sum^{k-1}_{s=r}\rho^{k-s-1}\gamma_s\varsigma(s)
\nonumber\\
\label{ne}&
\qquad+2L\delta_*\gamma_k\varsigma(k)+2\delta_*L\gamma_k\sum^2_{\ell=1}\frac{1}{n_\ell}\sum^{n_\ell}_{i=1}e_{i\ell }(k)\nonumber\\
&\qquad+2CL\delta_*\sum^{k-1}_{s=r}\gamma_s\rho^{k-s-1}\sum^2_{\ell=1}\sum^{n_\ell}_{i=1}e_{i\ell }(s)\\
&:=\sum^2_{\ell=1}\frac{1}{n_\ell}\textbf{1}'\Phi^\ell(k,r)\psi^\ell(r)\nonumber\\
\label{ex6}&\qquad-2\sum^{k-1}_{s=r}\gamma_s\sum^2_{\ell=1}\sum^{n_\ell}_{i=1}\vartheta^\ell_i(s)+\varrho(k,r),
\end{align}
where $\varsigma(s)=\max\{|x_i(s)-x^*|,i\in\mathcal{V}_1,|y_j(s)-y^*|,j\in\mathcal{V}_2\}$,
$\varrho(k,r)$ is the sum of all terms from the third one to the last one in
(\ref{ne}).

We next show $\lim_{r\rightarrow\infty}\sup_{k\geq r}\varrho(k,r)=0$.
First by Lemmas \ref{co}, \ref{con0} and Remark \ref{rem}, $\sum^{\infty}_{s=r}\gamma_s\sum^2_{\ell=1}\sum^{n_\ell}_{i=1}e_{i\ell }(s)<\infty$
and hence $\lim_{k\rightarrow\infty}\gamma_k\sum^2_{\ell=1}\sum^{n_\ell}_{i=1}e_{i\ell }(k)=0$.
It follows from $0<\rho<1$ that for each $k$,
$$
\sum^{k-1}_{s=r}\gamma_s\rho^{k-s-1}\sum^2_{\ell=1}\sum^{n_\ell}_{i=1}e_{i\ell }(s)
\leq \sum^{\infty}_{s=r}\gamma_s\sum^2_{\ell=1}\sum^{n_\ell}_{i=1}e_{i\ell }(s)<\infty.
$$
Moreover, by Lemma \ref{le2},
$\lim_{r\rightarrow\infty}\gamma_r\varsigma(r)=0$, which implies
$\lim_{r\rightarrow\infty}\sup_{k\geq r+1}\sum^{k-1}_{s=r}\rho^{k-s-1}\gamma_s\varsigma(s)=0$ along
with $\sum^{k-1}_{s=r}\rho^{k-s-1}\gamma_s\varsigma(s)\leq\frac{1}{1-\rho}\sup_{s\geq r}\gamma_s\varsigma(s)$.
From the preceding zero limit results, we have
$\lim_{r\rightarrow\infty}\sup_{k\geq r}\varrho(k,r)=0$. Then from (\ref{ex6})
$\sum^{\infty}_{s=r}\gamma_s\sum^2_{\ell=1}\sum^{n_\ell}_{i=1}\vartheta^\ell_i(s)<\infty$.
Clearly, from (\ref{sp}) $\sum^2_{\ell=1}\sum^{n_\ell}_{i=1}\vartheta^\ell_i(s)
=\Upsilon(s)\geq0$.
By the similar procedures to the proof of Theorem \ref{thm1}, we can show
that there is a subsequence $\{k_l\}$ such that
$\lim_{l\rightarrow\infty}\bar x(k_l)=x^*$,
$\lim_{l\rightarrow\infty}\bar y(k_l)=y^*$.

Now let us show $\lim_{k\rightarrow\infty}\sum^2_{\ell=1}\frac{1}{n_\ell}\textbf{1}'\psi^\ell(k)=0$.
First it follows from $\lim_{r\rightarrow\infty}\sup_{k\geq r}\varrho(k,r)=0$ that, for any $\varepsilon>0$, there is a sufficiently large $l_0$ such that
when $l\geq l_0$, $\sup_{k\geq k_l}\varrho(k,k_l)\leq \varepsilon$. Moreover,
since the consensus is achieved within the two subnetworks, $l_0$
can be selected sufficiently large such that $|x_i(k_{l_0})-x^*|\leq\varepsilon$
and $|y_i(k_{l_0})-y^*|\leq\varepsilon$ for each $i$.
With (\ref{ex6}), we have that, for each $k\geq k_l$,
\begin{align}
&\sum^2_{\ell=1}\frac{1}{n_\ell}\textbf{1}'\psi^\ell(k+1)
\leq\sum^2_{\ell=1}\frac{1}{n_\ell}\textbf{1}'\Phi^\ell(k,k_l)\psi^\ell(k_l)\nonumber\\
&\qquad\qquad+\sup_{k\geq k_l}\varrho(k,k_l)
\leq 2\varepsilon^2+\varepsilon,\nonumber
\end{align}
which implies $\lim_{k\rightarrow\infty}\sum^2_{\ell=1}\frac{1}{n_\ell}\textbf{1}'\psi^\ell(k)=0$.
Therefore, $\lim_{k\rightarrow\infty}x_i(k)=x^*$, $i\in \mathcal{V}_1$ and
$\lim_{k\rightarrow\infty}y_i(k)=y^*$, $i\in \mathcal{V}_2$.
Thus, the proof is completed. \hfill$\square$

\subsection{Proof of Theorem \ref{th4.5}}

 (i). In this case we design a dynamics for auxiliary states
$\alpha^i=(\alpha^i_1,\dots,\alpha^i_{n_1})'\in \mathbb{R}^{n_1}$ for
$i\in\mathcal{V}_1$ and $\beta^i=(\beta^i_1,\dots,\beta^i_{n_2})'\in \mathbb{R}^{n_2}$ for
$i\in\mathcal{V}_2$ to estimate the respective desired stepsizes:
\begin{equation}
\label{de} \begin{cases} \alpha^i(k+1)=\sum_{j\in\mathcal{N}^1_i(k)}a_{ij}(k)\alpha^j(k),\;k\geq0,\\
\beta^i(k+1)=\sum_{j\in\mathcal{N}^2_i(k)}a_{ij}(k)\beta^j(k),\;k\geq0\end{cases}
\end{equation}
with the initial value $\alpha^i_i(0)=1$, $\alpha^i_j(0)=0,\forall j\neq i$;
$\beta^i_i(0)=1$, $\beta^i_j(0)=0,\forall j\neq i$.

Then for each $i$ and $k$, let $\hat \alpha^i_k=\alpha^i_i(k)$,
$\hat \beta^i_k=\beta^i_i(k)$. Clearly, (\ref{conn}) holds.

First by {\bf A3} (i) and algorithm (\ref{de}), $\alpha^i_i(k)\geq \eta^k>0$ and $\beta^i_i(k)\geq
\eta^k>0$ for each $k$, which guarantees that the stepsize selection rule
(\ref{desi3}) is well-defined. Let
$\phi^\ell=(\phi^\ell_1,\dots,\phi^\ell_{n_\ell})'$ be the common
left eigenvector of $A_\ell(r),r\geq 0$ associated with eigenvalue one, where $\sum^{n_\ell}_{i=1}\phi^\ell_i=1$.
According to Lemma \ref{l1},
$\lim_{r\rightarrow\infty}\Phi^\ell(r,k)=\lim_{r\rightarrow\infty}A_\ell(r)\cdots
A_\ell(k)=\textbf{1}(\phi^\ell)'$ for each $k$. As a result,
$\alpha^i_k=\phi^1_i$, $i=1,...,n_1$; $\beta^i_k=\phi^2_i$,
$i=1,...,n_2$ for all $k$.

Let $\theta(k)=((\alpha^1(k))',\dots,(\alpha^{n_1}(k))')'$.
From (\ref{de}) we have
$$\theta(k+1)=(A_1(k)\otimes I_{n_1})\theta(k)$$
and then $\lim_{k\rightarrow\infty}\theta(k)
=\lim_{k\rightarrow\infty}(\Phi^1(k,0)\otimes I_{n_1})\theta(0)=(\textbf{1}(\phi^1)'\otimes I_{n_1})\theta(0)=\textbf{1}\otimes \phi^1$.
Therefore, $\lim_{k\rightarrow\infty}\alpha^i_i(k)=\phi^1_i$ for $i\in \mathcal{V}_1$.
Similarly, $\lim_{k\rightarrow\infty}\beta^i_i(k)=\phi^2_i$ for $i\in \mathcal{V}_2$.
Since $\alpha^i_k=\phi^1_i$ and $\beta^i_k=\phi^2_i$ for all $k$, (\ref{co1}) holds.
Moreover, the above convergence is achieved with a geometric rate by Lemma \ref{lem1}.
Without loss of generality, suppose $|\alpha^i_i(k)-\phi^1_i|\leq \bar C\bar \rho^k$
and $|\beta^i_i(k)-\phi^2_i|\leq \bar C\bar \rho^k$ for some $\bar C>0$, $0<\bar \rho<1$,
and all $i,k$.

The only difference between the models in Theorem \ref{th4.4} and
the current one is that the terms $\alpha^i_k$ and $\beta^i_k$
(equal to $\phi^1_i$ and $\phi^2_i$ in case (i), respectively) in
stepsize selection rule (\ref{desi2}) are replaced with $\hat
\alpha^i_k$ and $\hat \beta^i_k$ (equal to $\alpha^i_i(k)$ and
$\beta^i_i(k)$, respectively) in stepsize selection rule
(\ref{desi3}). We can find that all lemmas involved in the proof of
Theorem \ref{th4.4} still hold under the new stepsize selection rule
(\ref{desi3}). Moreover, all the analysis is almost the same as that
in Theorem \ref{th4.4} except that the new stepsize selection rule
will yield an error term (denoted as $\varpi^\ell(k,r)$) on the
right-hand side of (\ref{ex3}). In fact,
$$
\varpi^\ell(k,r)=2\sum^{k-1}_{s=r}\gamma_s\Phi^\ell(k,s+1)\varpi^\ell_s\vartheta^\ell(s)+
2\gamma_k\varpi^\ell_k\vartheta^\ell(k),
$$
where $\varpi^1_s=\mbox{diag}\big\{\frac{1}{\phi^1_1}-\frac{1}{\alpha^1_1(s)},\dots,\frac{1}{\phi^1_{n_1}}-\frac{1}{\alpha^{n_1}_{n_1}(s)}\big\}$,
$\varpi^2_s=\mbox{diag}\big\{\frac{1}{\phi^2_1}-\frac{1}{\beta^1_1(s)},\dots,\frac{1}{\phi^2_{n_2}}-\frac{1}{\beta^{n_2}_{n_2}(s)}\big\}$.
Moreover, since $\lim_{s\rightarrow\infty}\alpha^i_i(s)=\phi^1_i$,
$\alpha^i_i(s)\geq \phi^1_i/2\geq \eta^{(n_1-1)T_1}/2$,
\begin{align}
\Big|\frac{1}{\alpha^i_i(s)}-\frac{1}{\phi^1_i}\Big|=
\Big|\frac{\alpha^i_i(s)-\phi^1_i}{\alpha^i_i(s)\phi^1_i}\Big|
&\leq\frac{2|\alpha^i_i(s)-\phi^1_i|}{(\eta^{(n_1-1)T_1})^2}\leq\frac{2\bar C\bar \rho^s}{\eta^{2(n_1-1)T_1}}\nonumber
\end{align}
for a sufficiently large $s$. Analogously,
$\big|\frac{1}{\beta^i_i(s)}-\frac{1}{\phi^2_i}\big|\leq
\frac{2\bar C\bar \rho^s}{\eta^{2(n_2-1)T_2}}$. Then for a
sufficiently large $r$ and any $k\geq r+1$,
\begin{align}\label{l2}
&\Big|\sum^2_{\ell=1}\frac{1}{n_\ell}\textbf{1}'\varpi^\ell(k,r)\Big|\nonumber\\
&\leq4\bar CL\varepsilon_1\sum^{k-1}_{s=r}\gamma_s\bar \rho^s\max_{i,j}\{|x_i(s)-x^*|,|y_j(s)-y^*|\}\nonumber\\
&\leq4\bar CL\varepsilon_1\varepsilon_2\sum^{k-1}_{s=r}\bar \rho^s\leq4\bar CL\varepsilon_1\varepsilon_2\bar \rho^r/(1-\bar \rho),
\end{align}
where $\varepsilon_1=\max\{1/\eta^{2(n_1-1)T_1},1/\eta^{2(n_2-1)T_2}\},$
$\varepsilon_2=\sup_s\{\gamma_s\max_{i,j}\{|x_i(s)-x^*|,|y_j(s)-y^*|\}\}<\infty$
due to $\lim_{s\rightarrow\infty}\gamma_s\max_{i,j}\{|x_i(s)-x^*|,|y_j(s)-y^*|\}=0$
by Lemma \ref{le2}.
From the proof of Theorem \ref{th4.4}, we can find that
the relation (\ref{l2}) makes all the arguments hold
and then a Nash equilibrium is achieved for case (i).

(ii). Here we design a dynamics for the auxiliary states
$\alpha^{(\nu)i}=(\alpha^{(\nu)i}_1,\dots,\alpha^{(\nu)i}_{n_1})'$,
$\nu=0,...,p^1-1$ for $i\in\mathcal{V}_1$ and $\beta^{(\nu)i}=(\beta^{(\nu)i}_1,\dots,\beta^{(\nu)i}_{n_2})'$,
$\nu=0,...,p^2-1$ for $i\in\mathcal{V}_2$ to estimate the respective
desired stepsizes:
\begin{equation}
\label{de2}
\begin{cases}\alpha^{(\nu)i}(s+1)=\sum_{j\in\mathcal{N}^1_i(s)}a_{ij}(s)\alpha^{(\nu)j}(s),\\
\beta^{(\nu)i}(s+1)=\sum_{j\in\mathcal{N}^2_i(s)}a_{ij}(s)\beta^{(\nu)j}(s),
\end{cases}s\geq \nu+1
\end{equation}
with the initial value $\alpha^{(\nu)i}_i(\nu+1)=1$, $\alpha^{(\nu)i}_j(\nu+1)=0, j\neq i$;
$\beta^{(\nu)i}_i(\nu+1)=1$, $\beta^{(\nu)i}_j(\nu+1)=0, j\neq i$.

Then, for each $r\geq 0$, let $\hat
\alpha^i_{rp^1+\nu}=\alpha^{(\nu)i}_i(rp^1+\nu)$ for $i\in
\mathcal{V}_1$, $\nu=0,...,p^1-1$; let $\hat
\beta^i_{rp^2+\nu}=\beta^{(\nu)i}_i(rp^2+\nu)$ for $i\in
\mathcal{V}_2$, $\nu=0,...,p^2-1$.

Note that {\bf A2} implies that the union
graphs $\bigcup^{p^\ell-1}_{s=0}\mathcal{G}_{A^s_\ell},\ell=1,2$ are
strongly connected.
Let $\phi^{\ell(0)}$ be the Perron vector of $\lim_{r\rightarrow\infty}\Phi^\ell(rp^\ell-1,0)$,
i.e., $\lim_{r\rightarrow\infty}\Phi^\ell(rp^\ell-1,0)
=\lim_{r\rightarrow\infty}(A^{p^\ell-1}_\ell\cdots A^0_\ell)^r=\textbf{1}(\phi^{\ell(0)})'$.
Then for $\nu=1,...,p^\ell-1$,
\begin{align}\label{add}
&\lim_{r\rightarrow\infty}\Phi^\ell(rp^\ell+\nu-1,\nu)\nonumber\\
&=\lim_{r\rightarrow\infty}(A^{\nu-1}_\ell\cdots A^0_\ell A^{p^\ell-1}_\ell\cdots A^{\nu+1}_\ell A^\nu_\ell)^r\nonumber\\
&=\lim_{r\rightarrow\infty}(A^{p^\ell-1}_\ell\cdots A^0_\ell)^rA^{p^\ell-1}_\ell\cdots A^{\nu+1}_\ell A^\nu_\ell\nonumber\\
&=\textbf{1}(\phi^{\ell(0)})'A^{p^\ell-1}_\ell\cdots A^{\nu+1}_\ell A^\nu_\ell\; :=\textbf{1}(\phi^{\ell (\nu)})'.
\end{align}
Consequently, for each $r\geq 0$, $\alpha^i_{rp^1+\nu}=\phi^{1(\nu+1)}_i$, $\nu=0,1,...,p^1-2$,
$\alpha^i_{rp^1+p^1-1}=\phi^{1(0)}_i$.
Moreover, from (\ref{de2}) and (\ref{add}) we obtain that for $\nu=0,1,...,p^1-1$,
\begin{align}
&\lim_{r\rightarrow\infty}\theta^{\nu}(r)\nonumber\\
&=\big(\lim_{r\rightarrow\infty}\Phi^1(r,\nu+1)\otimes I_{n_1}\big)\theta^{\nu}(\nu+1)\nonumber\\
&=\big(\lim_{r\rightarrow\infty}\Phi^1(r,0)A^{p^1-1}_\ell\cdots A^{\nu+1}_\ell\otimes I_{n_1}\big)\theta^{\nu}(\nu+1)\nonumber\\
&=\big(\textbf{1}(\phi^{1(\nu+1)})'\otimes I_{n_1}\big)\theta^{\nu}(\nu+1),\nonumber
\end{align}
where
$\theta^{\nu}=((\alpha^{(\nu)1})',\dots,(\alpha^{(\nu)n_1})')'$,
$\phi^{1(p^1)}=\phi^{1(0)}$.
Then $\lim_{r\rightarrow\infty}\alpha^{(\nu)i}_i(r)=\phi^{1(\nu+1)}_i$ for $i\in \mathcal{V}_1$.
Hence,
$$
\lim_{r\rightarrow\infty}\big(\hat \alpha^i_{rp^1+\nu}-\alpha^i_{rp^1+\nu}\big)=0,\nu=0,...,p^1-1.
$$
Analogously, we have
$\lim_{r\rightarrow\infty}(\hat \beta^i_{rp^2+\nu}-\beta^i_{rp^2+\nu})=0,\nu=0,...,p^2-1.$
Moreover, the above convergence is achieved with a geometric rate.
Similar to the proof of case (i), we can prove case (ii).
Thus, the conclusion follows.  \hfill$\square$

\section{Numerical Examples}

In this section, we provide examples to illustrate the obtained
results in both the balanced and unbalanced graph cases.

Consider a network of five agents, where $n_1=3,n_2=2,m_1=m_2=1$,
$X=Y=[-5,5]$,
$f_1(x,y)=x^2-(20-x^2)(y-1)^2,f_2(x,y)=|x-1|-|y|,f_3(x,y)=(x-1)^4-2y^2$
and $g_1(x,y)=(x-1)^4-|y|-\frac{5}{4}y^2-\frac{1}{2}(20-x^2)(y-1)^2$,
$g_2(x,y)=x^2+|x-1|-\frac{3}{4}y^2-\frac{1}{2}(20-x^2)(y-1)^2$. Notice that
$\sum^3_{i=1}f_i=\sum^2_{i=1}g_i$ and all objective functions are
strictly convex-concave on $[-5,5]\times [-5,5]$. The unique saddle point of the sum objective function
$g_1+g_2$ on $[-5,5]\times [-5,5]$ is $(0.6102,0.8844)$.

Take initial conditions
$x_1(0)=2,x_2(0)=-0.5,x_3(0)=-1.5$ and $y_1(0)=1,y_2(0)=0.5$.
When $\hat x_2(k)=1,$ we take $q_{12}(k)=1\in \partial_x f_2(1,
\breve{x}_2(k))=[-1,1]$; when $\hat y_1(k)=0,$ we take
$q_{21}(k)=-1+(20-\breve{y}^2_1(k))\in \partial_y
g_1(\breve{y}_1(k),0)=\big\{r+(20-\breve{y}^2_1(k))|-1\leq r\leq
1\big\}$. Let $\gamma_k=1/(k+50)$, $k\geq0$, which satisfies {\bf A5}.

We discuss three examples. The first example is given for verifying the
 convergence of the proposed algorithm with homogeneous stepsizes in the case of weight-balanced graphs, while the second one is for the convergence with the stepsizes provided in the existence theorem
 in the case of weight-unbalanced graphs.  The third example demonstrates the efficiency
 of the proposed adaptive learning strategy for periodical switching unbalanced graphs.

\begin{exa}\label{exa1}
The communication graph is switching periodically over the two graphs $\mathcal{G}^e,\mathcal{G}^0$ given in Fig. 2,
where $\mathcal{G}(2k)=\mathcal{G}^e$, $\mathcal{G}(2k+1)=\mathcal{G}^o$, $k\geq0$.
\begin{figure}[!htbp]
\centering
\includegraphics[width=3.2in]{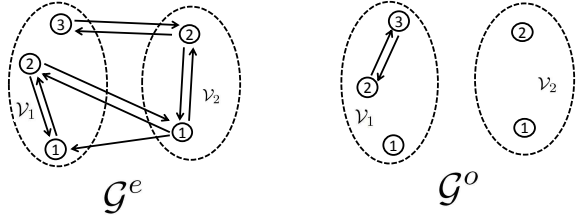}
\caption{Two possible communication  graphs in Example \ref{exa1}}
\end{figure}
Denote by $\mathcal{G}^e_1$ and $\mathcal{G}^e_2$ the two subgraphs of $\mathcal{G}^e$
describing the communications within the two subnetworks. Similarly,
the subgraphs of $\mathcal{G}^o$ are denoted as $\mathcal{G}^o_1$ and $\mathcal{G}^o_2$.
Here the adjacency matrices of $\mathcal{G}^e_1$, $\mathcal{G}^e_2$ and $\mathcal{G}^o_1$ are as follows:
$$
A_1(2k)=\left(
      \begin{array}{ccc}
        0.6 & 0.4 & 0 \\
        0.4 & 0.6 & 0 \\
        0 & 0 & 1 \\
      \end{array}
    \right),\;
    A_2(2k)=\left(
      \begin{array}{cc}
        0.9 & 0.1 \\
        0.1 & 0.9 \\
      \end{array}
    \right),
$$
$$
A_1(2k+1)=\left(
      \begin{array}{ccc}
        1 & 0 & 0 \\
        0 & 0.7 & 0.3 \\
        0 & 0.3 & 0.7 \\
      \end{array}
    \right).
$$

Clearly, with the above adjacency matrices, the three digraphs
$\mathcal{G}^e_1$, $\mathcal{G}^e_2$ and $\mathcal{G}^o_1$ are
weight-balanced.
Let the stepsize be $\alpha_{i,k}=\beta_{j,k}=\gamma_k$ for all $i,j$ and $k\geq0$.
Fig. 3 shows that the agents converge to the unique Nash
equilibrium $(x^*,y^*)=(0.6102,0.8844)$.

\begin{figure}[!htbp]
\centering
\includegraphics[width=2.8in]{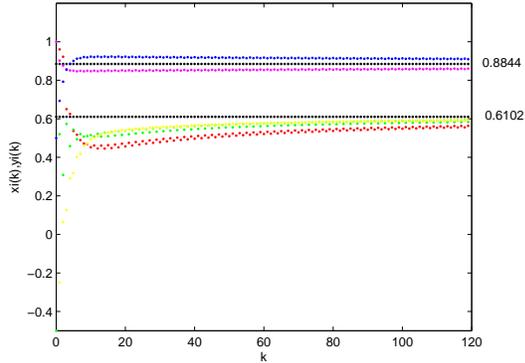}
\caption{The Nash equilibrium is achieved (i.e., $x_i\to x^*$ and
$y_i\to y^*$) for the time-varying weight-balanced digraphs with
homogeneous stepsizes.}
\end{figure}
\end{exa}

\begin{exa}\label{exa2}
Consider the same switching graphs given in Example \ref{exa1}
except that a new arc $(2,3)$ is added in $\mathcal{G}^e_1$.
The new graph is still denoted as $\mathcal{G}^e_1$ for simplicity.
Here the adjacency matrices of the
three digraphs $\mathcal{G}^e_1$, $\mathcal{G}^e_2$ and $\mathcal{G}^o_1$ are given by
$$
A_1(2k)=\left(
      \begin{array}{ccc}
        0.8 & 0.2 & 0 \\
        0.7 & 0.3 & 0 \\
        0 & 0.6 & 0.4 \\
      \end{array}
    \right),
A_2(2k)=\left(
      \begin{array}{cc}
        0.9 & 0.1 \\
        0.8 & 0.2 \\
      \end{array}
    \right),\;
$$
$$
A_1(2k+1)=\left(
      \begin{array}{ccc}
        1 & 0 & 0 \\
        0 & 0.3 & 0.7 \\
        0 & 0.4 & 0.6 \\
      \end{array}
    \right).
$$
In this case, $\mathcal{G}^e_1$, $\mathcal{G}^e_2$ and $\mathcal{G}^o_1$ are weight-unbalanced with
$(\alpha^1_{2k},\alpha^2_{2k},\alpha^{3}_{2k})=(0.5336,0.1525,0.3139)$, $(\alpha^1_{2k+1},\alpha^2_{2k+1},\alpha^{3}_{2k+1})=(0.5336,0.3408,0.1256)$
and $(\beta^1_k,\beta^2_k)=(0.8889,0.1111)$, $\forall k\geq0$.
We design the heterogeneous stepsizes as follows:
$\alpha_{i,2k}=\frac{1}{\alpha^i_{1}}\gamma_{2k},\;\alpha_{i,2k+1}=\frac{1}{\alpha^i_{0}}\gamma_{2k+1},i=1,2,3;\;
\beta_{i,k}=\frac{1}{\beta^i_{0}}\gamma_k,i=1,2$.
Fig. 4 shows that the agents converge to the unique Nash
equilibrium with those heterogeneous stepsizes.
\begin{figure}[!htbp]
\centering
\includegraphics[width=2.8in]{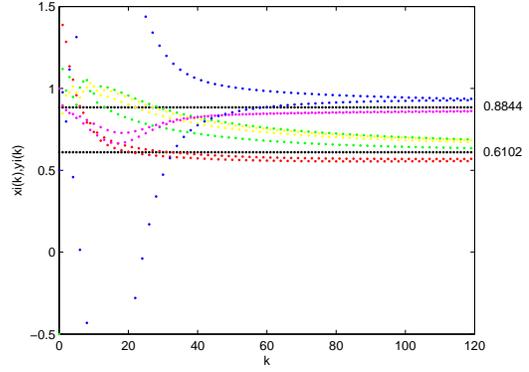}
\caption{The Nash equilibrium is achieved for weight-unbalanced
digraphs with heterogenous stepsizes.}
\end{figure}
\end{exa}

\begin{exa}
Here we verify the result obtained in Theorem \ref{th4.5} (ii).
Consider Example \ref{exa2}, where $p^1=p^2=2$.   Design adaptive stepsize algorithms:
for $\nu=0,1$,
$$
\theta^\nu(r)=(A_1(r)\cdots A_1(\nu+1)\otimes I_3)\theta^\nu(\nu+1),\;r\geq \nu+1,
$$
where $\theta^{\nu}(r)=((\alpha^{(\nu)1}(r))',(\alpha^{(\nu)2}(r))',(\alpha^{(\nu)3}(r))')'$,
$\theta^\nu(\nu+1)=(1,0,0,0,1,0,0,0,1)'$;
for $\nu=0,1$,
$$
\vartheta^\nu(r)=(A_2(r)\cdots A_2(\nu+1)\otimes I_2)\vartheta^\nu(\nu+1),\;r\geq \nu+1,
$$
where $\vartheta^{\nu}(r)=((\beta^{(\nu)1}(r))',(\beta^{(\nu)2}(r))')'$,
$\theta^\nu(\nu+1)=(1,0,0,1)'$.

 Let $\hat \alpha^i_{2k}=\alpha^{(0)i}_i(2k)$,
$\hat \alpha^i_{2k+1}=\alpha^{(1)i}_i(2k+1)$, $\hat \beta^i_{2k}=\beta^{(0)i}_i(2k)$,
$\hat \beta^i_{2k+1}=\beta^{(1)i}_i(2k+1)$ and
\begin{align}
&\alpha_{i,2k}=\frac{1}{\hat \alpha^i_{2k}}\gamma_{2k},\;\;
\alpha_{i,2k+1}=\frac{1}{\hat \alpha^i_{2k+1}}\gamma_{2k+1},\;i=1,2,3,\nonumber\\
&\beta_{i,2k}=\frac{1}{\hat \beta^i_{2k}}\gamma_{2k},\;\;
\beta_{i,2k+1}=\frac{1}{\hat \beta^i_{2k+1}}\gamma_{2k+1},\;i=1,2.
\nonumber
\end{align}
Fig. 5 shows that the agents converge to the unique Nash
equilibrium under the above designed adaptive stepsizes.
\begin{figure}[!htbp]
\centering
\includegraphics[width=2.8in]{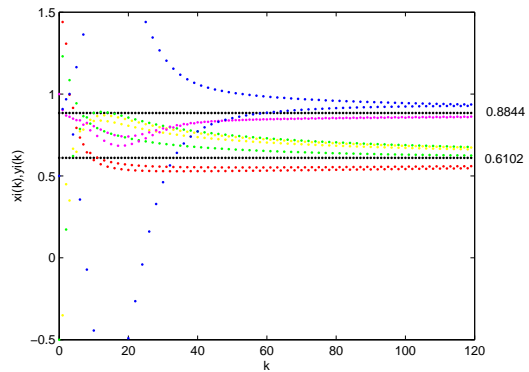}
\caption{The Nash equilibrium is achieved for weight-unbalanced digraphs by adaptive stepsizes.}
\end{figure}
\end{exa}

\section{Conclusions}

A subgradient-based distributed algorithm was proposed to solve a
Nash equilibrium computation problem as a zero-sum
game with switching communication graphs.  Sufficient conditions were provided to achieve a Nash
equilibrium for switching weight-balanced digraphs by an algorithm with
homogenous stepsizes.  In the case of weight-unbalanced graphs, it was demonstrated
that the algorithm with homogeneous stepsizes might fail to reach a
Nash equilibrium.  Then the existence of heterogeneous stepsizes to achieve a Nash
equilibrium was established.   Furthermore,
adaptive algorithms were designed to update the hoterogeneous stepsizes for the Nash
equilibrium computation in two special cases.

\end{document}